\documentclass[review]{siamart1116}

\usepackage{graphicx,graphics}
\usepackage{amsmath,amsfonts,mathrsfs,amssymb,dsfont}
\usepackage{booktabs}
\usepackage{cite}
\usepackage{sidecap}
\sidecaptionvpos{figure}{l}
\usepackage{bbm}
\usepackage{color}
\usepackage{arydshln}
\usepackage{upgreek}
\nolinenumbers

\usepackage{lipsum}
\usepackage{amsfonts}
\usepackage{graphicx}
\usepackage{epstopdf}
\usepackage{algorithmic}
\ifpdf
  \DeclareGraphicsExtensions{.eps,.pdf,.png,.jpg}
\else
  \DeclareGraphicsExtensions{.eps}
\fi

\numberwithin{theorem}{section}

\newcommand{\TheTitle}{Moment analysis of linear time-varying dynamical systems with renewal transitions} 
\newcommand{\TheShortTitle}{Moment analysis of linear time-varying TTSHS} 
\newcommand{\TheAuthors}{Mohammad Soltani, and Abhyudai Singh}

\headers{\TheShortTitle}{\TheAuthors}

\title{{\TheTitle}}

\author{Mohammad Soltani \thanks{ Department of Electrical and Computer Engineering, University of Delaware, Newark, DE USA 19716 (\email{msoltani@udel.edu}).}
  \and
  Abhyudai Singh\thanks{Department of Electrical and Computer Engineering, Biomedical Engineering, Mathematical Sciences, Center for Bioinformatics and Computational Biology, University of Delaware, Newark, DE USA 19716 (\email{absingh@udel.edu}).} 
}
\usepackage{amsopn}

\ifpdf
\hypersetup{
  pdftitle={\TheTitle},
  pdfauthor={\TheAuthors}
}
\fi

\begin{document}

\maketitle

\begin{abstract}
Stochastic dynamics of several systems can be modeled via piecewise deterministic time evolution of the state, interspersed by random discrete events. Within this general class of systems, we consider time-triggered stochastic hybrid systems (TTSHS), where the state evolves continuously according to a linear time-varying dynamical system. Discrete events occur based on an underlying renewal process (timer), and the intervals between successive events follow an arbitrary continuous probability density function. Moreover, whenever the event occurs, the state is reset based on a linear affine transformation that allows for the inclusion of state-dependent and independent noise terms. Our key contribution is derivation of necessary and sufficient conditions for the stability of statistical moments, along with exact analytical expressions for the steady-state moments. These results are illustrated on an example from cell biology, where deterministic synthesis and decay of a gene product (RNA or protein) is coupled to random timing of cell-division events. As experimentally observed, cell-division events occur based on an internal timer that measures the time elapsed since the start of cell cycle (i.e., last event). Upon division, the gene product level is halved, together with a state-dependent noise term that arises due to randomness in the partitioning of molecules between two daughter cells.  We show that the TTSHS framework is conveniently suited to capture the time evolution of gene product levels, and derive unique formulas connecting its mean and variance to underlying model parameters and noise mechanisms. Systematic analysis of the formulas reveal counterintuitive insights, such as, if the partitioning noise is large then making the timing of cell division more random reduces noise in gene product levels. In summary, theory developed here provides novel tools for characterizing moments in an important class of stochastic dynamical systems that arises naturally in diverse application areas.
\end{abstract}



\section{Introduction}
We study a class of stochastic systems that couple continuous linear dynamics with random discrete events that occur based on an underlying renewal process. Such systems
have been referred to in literature as time-triggered stochastic hybrid systems (TTSHS) \cite{dav84,cos90,dav93,cod08}, and are an important sub-class of piecewise-deterministic Markov processes (PDMP) \cite{flj92,fgd00,cfm08,ddz15} with applications in different disciplines. For example, TTSHS have been shown to arise ubiquitously in networked control systems, where a dynamical system is controlled over a noisy communication network, and signals are received at discrete random times \cite{ahs12,hes14,ahs13a,gad14,hjt12,ank16,sos17b,ant10,aah17}. Other TTSHS applications include modeling disturbances in nanosensors \cite{sos16c}, capturing stochastic effects in cellular biochemical processes \cite{ans14,sva15, sos16,dsp15}, and neuroscience \cite{sin17b}.

Previously, we studied a sub-class of TTSHS where the continuous dynamics was modeled by a linear time-invariant system, and the time intervals between successive discrete events was restricted to follow a phase-type distribution (i.e., mixture and/or sum of exponential random variables) \cite{sos16c}.  For such systems, statistical moments of the state space can be computed exactly by numerically solving a system of differential equations \cite{sos16c}. Building up on this prior work, here we generalize the results in several new directions:\\
\begin{itemize}
\item Allow continuous dynamics to be a time-varying linear system. \\
\item  Time intervals between events  follow an arbitrary positively-valued and continuous probability density function (pdf). \\
\item  Provide explicit condition for the existence and convergence of statistical moments, together with their exact closed-form formulas.\\
\item  Use  TTSHS to study the fundamental process of gene expression inside cells, where production/decay of a protein is coupled to random cell-division events. \\
\end{itemize}
We start by introducing the notation used throughout the paper, followed by a mathematical description of TTSHS. Before presenting the results on TTSHS with time-varying
dynamics, we first consider the simpler case of linear time-invariant systems. 

\textit{Notation}: 
The set of real number is denoted by $\mathbb{R}$. Constant vectors are indicated by a hat, e.g. $\hat{a}$, and matrices are denoted by capital letters. Further, transpose of a matrix $A$ is given by $A^\top$ and the n-dimensional identity matrix is denoted by $I_n$. We show zero vectors and matrices with the same notation, e.g. $A=\hat{a} = 0$. Random variables are indicated by bold letters. The expected value of a random variable $\boldsymbol x $ is denoted by $\langle \boldsymbol x \rangle$ and the expected value in steady-state is denoted by $\overline{\langle  { \boldsymbol x}  \rangle } \equiv\lim_{t\to\infty}\langle \boldsymbol x \rangle$. Finally, the conditional value $\boldsymbol x$ given another random variable $\boldsymbol y$ is denoted $\boldsymbol x\vert_{\boldsymbol y} $. 

\section{Linear time-invariant TTSHS}
The state of the system $ {\boldsymbol x} \in \mathbb{R}^{n \times 1}$ evolves as per the  
following ordinary differential equation (ODE)
\begin{equation}
\frac{d {\boldsymbol x}}{dt}= \hat{a}+A {\boldsymbol x}(t), \label{dynamics0000}
\end{equation}
for a given constant vector $\hat{a} \in \mathbb{R}^{n\times 1}$,  and matrix $A \in \mathbb{R}^{n\times n}$. 
Random events are assumed to occur at times $\boldsymbol{t}_s, \ s\in \{1,2,\ldots\}$, and the time interval between events
\begin{align}
\boldsymbol \tau_s  \equiv   \boldsymbol t_s - \boldsymbol  t_{s-1}\label{pdf of T}
\end{align}
is an independent and identically distributed (iid) random variable that follows a continuous positively-valued pdf $f$. Throughout the paper we assume a finite
mean time interval $\langle \boldsymbol \tau_s \rangle <\infty$, but higher-order moments  of $\boldsymbol \tau_s$ can be infinite allowing for heavy-tailed timing distributions.

Whenever the events occur the state is reset as
\begin{equation}
 {\boldsymbol x}(\boldsymbol t_s^-)\mapsto  {\boldsymbol x}(\boldsymbol  t_s^+), \label{reset}
\end{equation}
where $ {\boldsymbol x}(\boldsymbol t_s^-)$ and $ {\boldsymbol x}(\boldsymbol t_s^+)$ denote the state of the TTSHS just before and after the event, respectively.  We assume $ {\boldsymbol x}(\boldsymbol  t_s^+)$ to be a random variable, whose average value is related to its value just before the event by a linear affine map 
\begin{align}
&\langle  {\boldsymbol{x}} (\boldsymbol{t}_s^+)\rangle=  J {\boldsymbol{x}}(\boldsymbol{t}_s^-)+ \hat{r},\label{conditional x0}
\end{align}
where $J\in \mathbb{R}^{n\times n}$ and $\hat{r} \in \mathbb{R}^{ n \times 1}$ are a constant matrix and vector, receptively.  Furthermore, the covariance matrix of $ {\boldsymbol x}(\boldsymbol  t_s^+)$
is defined by
\begin{equation}
\begin{aligned}
\langle  {\boldsymbol{x}}(\boldsymbol{t}_s^+)  {\boldsymbol{x}}^\top (\boldsymbol{t}_s^+) \rangle & - \langle  {\boldsymbol{x}} (\boldsymbol{t}_s^+)\rangle\langle  {\boldsymbol{x}} (\boldsymbol{t}_s^+)\rangle^\top= \\ &	\label{conditional x20}
Q   {\boldsymbol{x}}(\boldsymbol{t}_s^-) {\boldsymbol{x}}^\top(\boldsymbol{t}_s^-) Q^\top+ B    {\boldsymbol{x}} (\boldsymbol{t}_s^-) \hat{c}^\top+ \hat{c}  {\boldsymbol{x}}^\top  (\boldsymbol{t}_s^-)B^\top+ D.
\end{aligned}
\end{equation}
Here $ Q \in \mathbb{R}^{n \times n}$ and $ B \in \mathbb{R}^{n \times n}$ are constant matrices, and $ \hat{c}\in \mathbb{R}^{n \times 1}$ is a constant vector. Moreover $ D \in \mathbb{R}^{ n \times n}$ is a constant symmetric positive semi-definite matrix. Intuitively, \eqref{conditional x20} formalizes the noise added to the state during the reset (event), 
with $Q=B=D=\hat{c}=0$ implying that ${\boldsymbol{x}} (\boldsymbol{t}_s^+)$ is simply a deterministic linear function of ${\boldsymbol{x}} (\boldsymbol{t}_s^-)$. A constant state-independent noise can be incorporated through 
a nonzero matrix $D$ with $Q=B=\hat{c}=0$. The generality of \eqref{conditional x20} allows for state-dependent noise terms that can potentially be quadratic (nonzero $Q$) or linear (nonzero $B$ and $\hat{c}$) functions of the state, and we will see an example of it later in the manuscript. 

\begin{SCfigure}[][!btht]
	\centering
	{\includegraphics[width=0.5\columnwidth]{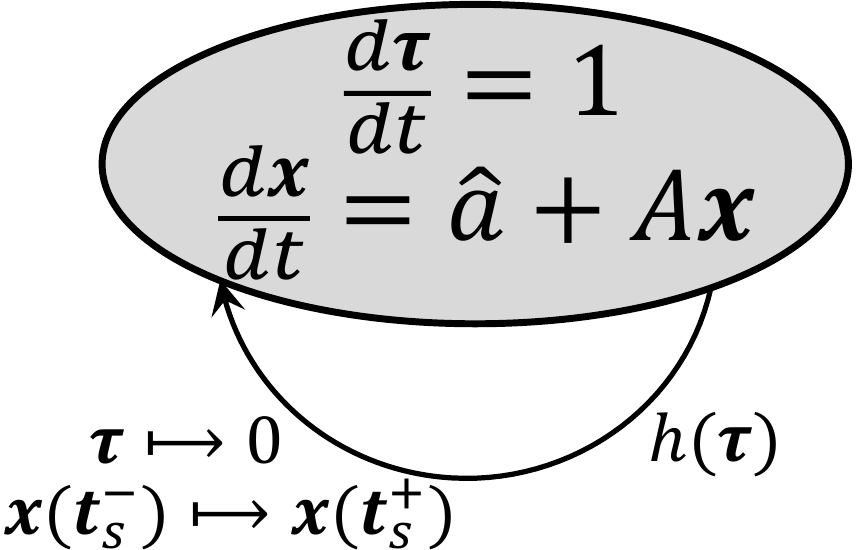}}
	\caption{{\bf Schematic of TTSHS with continuous dynamics described by a linear time-invariant system}. As the state evolves according to a linear system, events occur at discrete times that change the state of the system according to \eqref{reset}. The timing of events is controlled by renewal transitions defined through a timer $\boldsymbol \tau$ that linearly increases over time in between events, and is reset to zero each time an event occurs. Choosing the event arrival rate $h(\boldsymbol \tau)$ based on \eqref{hr} ensures that the time interval between events is iid with probability distribution $f$.}
	\label{fig:model}
\end{SCfigure}

\section{Statistical analysis of linear time-invariant TTSHS}
A convenient approach to implement the  TTSHS represented by \eqref{dynamics0000}-\eqref{conditional x20} is via a timer $\boldsymbol \tau$ that measures the time elapsed since the last event (Fig. 1). The timer increases between events, and resets to zero whenever the events occur. Let the 
probability that an event occurs in the next infinitesimal time $(t,t+dt]$ be $h(\boldsymbol \tau)dt$, where
\begin{equation} \label{hr}
 h( \tau ) \equiv   \frac{f(\tau)}{1-\int_{y=0}^{\tau}f(y)dy}
\end{equation}
is the event arrival rate (hazard rate). Then, $\boldsymbol \tau_s$ follows the continuous positively-valued pdf 
\begin{equation}
\boldsymbol \tau_s \sim  f(\tau) = h(\tau )  e^{-\int_0^{\tau} h(y) d y}
\end{equation}
\cite{Ross20109,fin08,ehp00}, and at steady-state, the timer follows the continuous positively-valued pdf
\begin{equation}
\boldsymbol \tau \sim  p(\tau) = \frac{1}{\langle \boldsymbol \tau_s\rangle} e^{-\int_0^{\tau} h(y) d y} \label{prob. tau}
\end{equation}
\cite{vss16}. As a simple example, a constant (timer independent) hazard rate $ h(\tau )=1/\langle \boldsymbol \tau_s\rangle$ leads to exponentially-distributed $ \boldsymbol \tau_s$.
Similarly, a monomial function
\begin{equation} \label{weibull}
 h(\tau )=  \frac{k}{\lambda} \left(\frac{\tau}{\lambda}\right)^{k-1},
\end{equation}
with positive constants $k$ and $\lambda$ results in a Weibull distribution for $\boldsymbol \tau_s$ with pdf
\begin{equation} 
f(\tau)= \frac{k}{\lambda} \left(\frac{\tau}{\lambda}\right)^{k-1}  e^{-(\tau/\lambda)^k} 
\end{equation}
and mean $\langle \boldsymbol \tau_s\rangle=\lambda \ \Gamma (1+1/k)$, where $\Gamma$ is the gamma function. Having defined the probability distributions of $\boldsymbol \tau_s$ and $\boldsymbol \tau$, we next summarize our main results  in theorems/corollaries, and refer the reader to the appendix for proofs.

\subsection{Mean of vector $ {\boldsymbol x}$}
In general, the expected value of $ {\boldsymbol x}$ depends on the entire distribution of $\boldsymbol \tau_s$, as shown below.
\begin{theorem} For the TTSHS \eqref{dynamics0000}-\eqref{conditional x20} the steady-state mean of $ {\boldsymbol x}$ exsists and is given by
	\begin{equation}
			\small
			\overline{\langle  { \boldsymbol x}  \rangle } 
	 =  \left\langle e^{ A\boldsymbol \tau} \right \rangle \left(I_n -J \left\langle e^{ A\boldsymbol \tau_s} \right \rangle   \right)^{-1}\left( J\left \langle  e^{ A\boldsymbol \tau_s} \int_0^{\boldsymbol \tau_s}  e^{ -Al} \hat{a} dl  \right \rangle + \hat{r} \right)   + \left \langle e^{A \boldsymbol \tau}  \int_0^{\boldsymbol \tau}  e^{ -Al} \hat{a} dl  \right \rangle
	\label{mean of x}	
	\end{equation} 
if and only if the expected value 
	\begin{equation}
 \left\langle  e^{ A\boldsymbol \tau _s} \right \rangle =  \int_0^{\infty}f(\tau) e^{A\tau }d \tau 
	\end{equation}
exists and all the eigenvalues of the matrix $J \left\langle  e^{ A\boldsymbol \tau _s} \right \rangle$ are inside the unit circle. 
\end{theorem} 
Please see Appendix A for a detailed proof. In this theorem, the vector
\begin{equation} \small
 \left \langle  e^{ A\boldsymbol \tau_s} \int_0^{\boldsymbol \tau_s}  e^{ -Al} \hat{a} dl  \right \rangle  =  \int_0^{\infty} f(\tau )\left( e^{ A\tau } \int_0^{\tau }  e^{ -Al} \hat{a} dl \right)  d \tau 
\end{equation}
is obtained by taking the expected value with respect to $\boldsymbol \tau_s$, and 
\begin{equation} \small
\left\langle  e^{ A\boldsymbol \tau } \right \rangle =  \int_0^{\infty}p(\tau) e^{A\tau }d \tau, \ \ \ \left \langle  e^{ A\boldsymbol \tau } \int_0^{\boldsymbol \tau }  e^{ -Al} \hat{a} dl  \right \rangle  =  \int_0^{\infty} p(\tau )\left( e^{ A\tau } \int_0^{\tau }  e^{ -Al} \hat{a} dl \right)  d \tau 
\end{equation}
is obtained by taking the expected value with respect to $\boldsymbol \tau$.  While Theorem 3.1 represents the most general result, we consider 
simplifications of \eqref{mean of x} in special cases. 
\begin{corollary}
If the  TTSHS \eqref{dynamics0000}-\eqref{conditional x20} satisfies Theorem 3.1 and the matrix $A$ is invertible, then 
\end{corollary}
\begin{equation}\label{mean x0}
\begin{aligned}
\overline{\langle  {\boldsymbol{x}}  \rangle} = & \frac{1}{\langle \boldsymbol \tau_s \rangle }\left(  I_n- \left\langle e^{ A  \boldsymbol  \tau_s}  \right \rangle \right)A^{-1}  \left(I_n -J\left\langle  e^{ A  \boldsymbol \tau_s} \right \rangle  \right)^{-1} 
\left( J\left( I_n - \left\langle e^{ A  \boldsymbol  \tau_s}  \right \rangle \right) A^{-1} \hat{a}  +\hat{r}  \right) \\ & 
- \frac{1}{\langle \boldsymbol \tau_s \rangle } \left(  I_n - \left\langle e^{ A  \boldsymbol  \tau_s}  \right \rangle \right)  A^{-2}  \hat{a} - A^{-1}  \hat{a}.
\end{aligned}
\end{equation} 
Thus, for an invertible matrix $A$, the steady-state expected value can directly be computed from the moment generating function $\left\langle  e^{ A  \boldsymbol \tau_s} \right \rangle$
(see Appendix B). Interestingly, there are some scenarios where knowing a few lower-order moments of $\boldsymbol \tau_s$ are sufficient to determine $\overline{\langle  {\boldsymbol{x}} \rangle}$ (see Appendix C).

\begin{corollary}
Consider the TTSHS \eqref{dynamics0000}-\eqref{conditional x20} with $A=0$, and all eigenvalues of the matrix $J$ are inside the unit circle, then
\begin{equation}
\begin{aligned}
& \overline{\langle  {\boldsymbol{x}} \rangle}=  \left(I_n -J  \right)^{-1}
\left( J \left \langle \boldsymbol \tau_s \right \rangle \hat{a} + \hat{r} \right) + \frac{\langle \boldsymbol \tau_s^2 \rangle}{2 \langle \boldsymbol \tau_s \rangle }\hat{a} . \label{mean of x A=0}
\end{aligned}
\end{equation} 	
only depends on the  first- and the second-order moments of $ \boldsymbol \tau_s$.
\end{corollary} 
We will revisit this corollary later on, as it is pertinent to the example of gene expression.

\subsection{Second-order moments of TTSHS}
In order to calculate the second-order moments, we start by deriving the dynamics of $ {\boldsymbol x}   {\boldsymbol x}^\top$ in between two successive events
\begin{equation} \label{non vector}
\frac{d\left(  {\boldsymbol x}   {\boldsymbol x}^\top\right)}{dt}=
\frac{d   {\boldsymbol x}}{dt} {\boldsymbol x}^\top+  {\boldsymbol x} \frac{d   {\boldsymbol x}^\top}{dt}= 
 A {\boldsymbol x}   {\boldsymbol x}^\top+  {\boldsymbol x}   {\boldsymbol x}^\top A^\top+ \hat{a}   {\boldsymbol x}^\top+   {\boldsymbol x} \hat{a}^\top.
\end{equation}
To proceed further, we introduce a new transformation named ``vectorization", i.e., a linear transformation that converts a matrix into a column vector. For instance 
\begin{equation}
A =     \left[\begin{array}{cc}
a_{11} & a_{12}\\ a_{21}  & a_{22}
	\end{array}\right] \Rightarrow {\rm vec}(A)= \left[\begin{array}{cccc}
a_{11} & a_{21} & a_{12} &  a_{22}
	\end{array}\right]^\top ,
\end{equation}
where ${\rm vec}()$ stands for the vectorization of a matrix. By putting all the columns of the matrix $ {\boldsymbol x}   {\boldsymbol x}^\top $ into one vector
 ${\rm vec}\left( {\boldsymbol x}   {\boldsymbol x}^\top\right)\in \mathbb{R}^{n^2 \times 1}$, \eqref{non vector} can be transformed as
\begin{equation} \label{veced}
\frac{d{\rm vec}\left( {\boldsymbol x}   {\boldsymbol x}^\top\right)}{dt} =  (I_n \otimes A + A\otimes I_n){\rm vec}\left({ {\boldsymbol x}   {\boldsymbol x}}^\top\right) +( I_n \otimes \hat{a} + \hat{a}  \otimes I_n) {\boldsymbol x} ,
\end{equation}
where $\otimes$ denotes the Kronecker product. Note that in transforming \eqref{non vector} to \eqref{veced} we used the fact that for three matrices $M_1$, $M_2$, and $M_3$ 
\begin{equation}\begin{aligned}
& {\rm vec}(M_1 M_2 M_3) = (M_3^\top \otimes M_1){\rm vec}(M_2) \\
& \hspace{25mm}\Rightarrow 
 \left \lbrace \begin{array}{l}
{\rm vec}( A {\boldsymbol x}   {\boldsymbol x}^\top) ={\rm vec}( A {\boldsymbol x}   {\boldsymbol x}^\top I_n ) =( I_n \otimes A) {\rm vec}(  {\boldsymbol x}   {\boldsymbol x}^\top), \\
{\rm vec}(  {\boldsymbol x}   {\boldsymbol x}^\top A^\top) ={\rm vec}(I_n  {\boldsymbol x}   {\boldsymbol x}^\top A^\top) = (A\otimes I_n) {\rm vec}(  {\boldsymbol x}   {\boldsymbol x}^\top), \\ 
{\rm vec}(\hat{a}   {\boldsymbol x}^\top) ={\rm vec}(\hat{a}   {\boldsymbol x}^\top I_n) =  ( I_n \otimes \hat{a} ) {\boldsymbol x} ,\\
{\rm vec}(   {\boldsymbol x} \hat{a}^\top)= {\rm vec}( I_n  {\boldsymbol x} \hat{a}^\top)= (\hat{a}  \otimes I_n) {\boldsymbol x} 
\end{array} \right. 
\end{aligned}
 \label{kronecker}
\end{equation}
\cite{mao13}. It turns out that if we define a vector  $ {\boldsymbol \mu}\equiv \left[ {\boldsymbol x}^\top \ \ \ {\rm vec}  \left(  {\boldsymbol x}   {\boldsymbol x}^\top \right)^\top\right]^\top\in \mathbb{R}^{(n+n^2) \times 1}$, its time evolution can also  be represented by a TTSHS, albeit a more complex one. More specifically,
\begin{equation}
\frac{d {\boldsymbol \mu} }{dt} =\hat{a}_\mu + A_\mu   {\boldsymbol \mu} , \label{mu dynamics}
\end{equation}
in between two successive events, where
\begin{equation}\label{Auau0}
\begin{aligned}
& 
A_\mu\equiv & \left[\begin{array}{c;{2pt/2pt}c}
A & 0\\ \hdashline[2pt/2pt] I_n \otimes \hat{a} + \hat{a}  \otimes I_n & I_n \otimes A + A\otimes I_n 
\end{array}\right],  
\hat{a}_\mu \equiv \left[\begin{array}{c}
\hat{a}\\ \hdashline[2pt/2pt] 0
\end{array}\right].
\end{aligned}
\end{equation}
Furthermore, whenever an event occurs,  $ {\boldsymbol\mu}$ is reset as
 \begin{equation}
 {\boldsymbol \mu} (\boldsymbol{t}_s^-)  \mapsto  {\boldsymbol \mu} (\boldsymbol{t}_s^+) , \label{reset 2}
 \end{equation}
where the expected value of ${\boldsymbol \mu} (\boldsymbol{t}_s^+) $ is given by (see Appendix D.1)
 \begin{subequations}\label{Jmu}
 	\begin{align}
&	\langle  {\boldsymbol \mu} (\boldsymbol{t}_s^+)  \rangle =   J_\mu   {\boldsymbol \mu}  (\boldsymbol{t}_s^-) + \hat{r}_\mu,\\
& J_\mu \equiv    \left[\begin{array}{c;{2pt/2pt}c}
J & 0\\ \hdashline[2pt/2pt] \begin{array}{c} B\otimes \hat{c}+J\otimes \hat{r} \\  +\hat{c}\otimes B+\hat{r}\otimes J\end{array}  &  J\otimes J+Q\otimes Q 
	\end{array}\right],
 \hat{r}_\mu \equiv  \left[\begin{array}{c}
\hat{r}\\ \hdashline[2pt/2pt]  {\rm vec }(D +\hat{r}\hat{r} ^\top ) 
 \end{array}\right]. 
 \end{align}
\end{subequations}
In summary, we have recast the stochastic dynamic of ${\boldsymbol \mu}$ as a TTSHS \eqref{mu dynamics} -\eqref{Jmu}, and a similar analysis as in Theorem 3.1 leads to the following result (see Appendix D.2).\begin{theorem}
Assuming the original TTSHS given by \eqref{dynamics0000}-\eqref{conditional x20} satisfies Theorem 3.1.  Then 
	\begin{equation}
	\begin{aligned}
\overline{\langle   {\boldsymbol \mu}   \rangle} = &  \left \langle e^{ A_\mu\boldsymbol \tau  } \int_0^{\boldsymbol \tau}   e^{ -A_\mu l} \hat{a}_\mu dl \right \rangle\\ & + \langle e^{A_\mu\boldsymbol \tau} \rangle   \left(I_{n^2+n} -J_\mu \left\langle e^{ A_\mu\boldsymbol \tau_s} \right \rangle   \right)^{-1}\left( J_\mu \left \langle  e^{ A_\mu \boldsymbol \tau_s} \int_0^{\boldsymbol \tau_s}  e^{ -A_\mu l} \hat{a}_\mu  dl  \right \rangle + \hat{r}_\mu  \right) 
\label{mu00}	\end{aligned} 
	\end{equation}
if and only if all the eigenvalues of the matrix $(J\otimes  J+Q\otimes Q) \left\langle  e^{ A\boldsymbol \tau_s} \otimes e^{ A\boldsymbol \tau_s} \right \rangle $ are inside the unit circle. 	
\end{theorem} 
Theorems 3.1 and 3.4 provide sufficient conditions for the existence of the first two moments of $\boldsymbol{x}$.

\textbf{Remark 1}: 
lf $A$ is a Hurwitz matrix (i.e., the deterministic continuous dynamics is by itself stable), $J$ is a diagonal positive definite matrix and all of its eigenvalues are inside the unit circle, then 
the steady-state mean of $\boldsymbol x$ exists. Moreover, if $Q$ is diagonal, $ J\otimes  J+Q\otimes Q$ is positive definite and all of its eigenvalues are inside the unit circle, then the second-order moments of $\boldsymbol x$ also exists (see Appendix E). Note that in these cases the first two moments of $\boldsymbol{x}$ remain bounded even though higher-order moments  of $\boldsymbol \tau_s$ may be unbounded.\\

The different corollaries of Theorem 3.1 that consider special cases can also be generalized to Theorem 3.4. For instance, if $A_\mu$ is invertible then similar to Corollary 3.2, the steady-state mean of vector $ {\boldsymbol \mu} $ takes the form
\begin{equation}\label{mean x20}
\begin{aligned}
&  \overline{ \langle   {\boldsymbol \mu} \rangle} =  \frac{1}{\langle \boldsymbol \tau_s \rangle }\left(  I_{n^2+n}- \left\langle e^{ A_\mu   \boldsymbol  \tau_s}  \right \rangle \right)A_\mu^{-1}  \left(I_{n^2+n} -J_\mu\left\langle  e^{ A_\mu   \boldsymbol \tau_s} \right \rangle  \right)^{-1} \times 
\\ & \left( J_\mu\left(  I_{n^2+n}- \left\langle e^{ A_\mu  \boldsymbol  \tau_s}  \right \rangle \right) A_\mu^{-1} \hat{a}  + \hat{r}_\mu  \right)
- \frac{1}{\langle \boldsymbol \tau_s \rangle } \left(  I_{n^2+n}- \left\langle e^{ A_\mu   \boldsymbol  \tau_s}  \right \rangle \right)  A_\mu^{-2}  \hat{a}_\mu - A_\mu^{-1}  \hat{a}_\mu.
\end{aligned}
\end{equation} 
Moreover, as an extension of Corollary 3.3, we show in Appendix F that when $A=0$, $\overline{\langle \boldsymbol x \boldsymbol x^\top \rangle}$ only depends on the first three moments of $\boldsymbol \tau_s$. 

Finally, we apply Theorem 3.4 to a  subclass of TTSHS where matrix $A$ is Hurwitz and  $Q=\hat{r}=0$,  $J=I_n$ in \eqref{conditional x0}-\eqref{conditional x20}, which corresponds to 
$\langle  {\boldsymbol{x}} (\boldsymbol{t}_s^+)\rangle= {\boldsymbol{x}}(\boldsymbol{t}_s^-)$ and
\begin{equation}
\begin{aligned}
 \langle  {\boldsymbol{x}}(\boldsymbol{t}_s^+)  {\boldsymbol{x}}^\top (\boldsymbol{t}_s^+) \rangle  - \langle  {\boldsymbol{x}} (\boldsymbol{t}_s^+)\rangle\langle  {\boldsymbol{x}} (\boldsymbol{t}_s^+)\rangle^\top= 
 B    {\boldsymbol{x}} (\boldsymbol{t}_s^-) \hat{c}^\top+ \hat{c}  {\boldsymbol{x}}^\top  (\boldsymbol{t}_s^-)B^\top+ D.
\end{aligned}
\end{equation}
Here discrete events do not affect the mean behaviour of the system but function to impart noise at random times.  We have previously studied these systems in the context of nanosensors, where gas molecules impinging on the sensor strike at random times and change the sensor velocity by adding a zero-mean noise term \cite{sos16c}. Using Corollary 3.2 and Theorem 3.4 for this subclass results in
$ \overline{ \langle  {\boldsymbol x} \rangle}=-A^{-1}\hat{a}$ and  
	\begin{equation} \label{noisy reset}
	\begin{aligned}
 {\rm vec} &	(\overline{\langle  {\boldsymbol{x}} {\boldsymbol{x}}^T \rangle })= -(I_n\otimes A+A\otimes I_n)^{-1} \times 
 \\  &\left( \left(\frac{1}{\langle \boldsymbol \tau_s\rangle} B\otimes \hat{c}+\frac{1}{\langle \boldsymbol \tau_s\rangle} \hat{c}\otimes B+I_n\otimes \hat{a}+\hat{a}\otimes I_n\right)\overline{\langle  {\boldsymbol{x}} \rangle} + \frac{1}{\langle \boldsymbol \tau_s\rangle} {\rm vec}(D)\right).
	\end{aligned}
	\end{equation}	
As expected, the steady-state mean is independent of $ \boldsymbol \tau_s$. Counterintuitively, the second-order moment only depends on the mean arrival times $\langle \boldsymbol \tau_s\rangle$, and making the timing of events more stochastic for fixed $\langle \boldsymbol \tau_s\rangle$ will not have any effect on the magnitude of random fluctuations in  $ {\boldsymbol x}$. We next illustrate the theory developed for TTSHS to the biological example of gene expression.

\section{Quantifying noise in gene expression via TTSHS}
The process of gene expression by which information encoded in DNA is used to synthesize gene products (RNAs and proteins) is fundamental to life. Measurements of gene product levels inside individual cells reveal a striking heterogeneity: the level of a gene product can vary considerably among cells of the same population, in spite of the fact that cells are genetically identical and are exposed to the same extracellular environment \cite{els02,bkc03,brb14,ngi06,bpm06,rao05}. Such cell-to-cell variation or expression noise critically impact functioning of intracellular circuits \cite{ele10,rao08,keb05,lps07}; drives seemingly identical cells to different fates \cite{lod08,arm98,Maamar27072007,Abranches01072014,wds08,nlp15}; and also implicated in emerging medical problems, such as, HIV latency \cite{siw09,wbt05,singh2012stochastic,rpa15}, cancer drug resistance \cite{sdt17}, and bacterial persistence \cite{bmc04,mcc13,kd_12,Rotem28062010,vsk08,kul05}. Thus, uncovering noise mechanisms that lead to cell-to-cell expression variation has tremendous implications for both biology and medicine. 

Considerable theoretical and experimental research over the last decade has primarily focused on characterizing stochasticity inherent in the different steps of gene expression  \cite{src10,sis13,moh15c,sbf15,srd12,cfx06,shs08,sih09b,otk02,drs12,rpt06,dss16}. Here we focus on a different mechanism: noise in the timing of cell cycle, or time taken by a newborn cell to complete its cell cycle and divide into two daughter cells. While much work coupling gene expression with cell cycle considers deterministic timing of division  \cite{schwabe_2014,huh11,kvw15,hup11}, data across organisms point to cell-cycle times following a non-exponential distribution that is often approximated by a lognormal or gamma distribution \cite{lambert_2015,tsukanov_2011,reshes_timing_2008,Roeder:2010vb,Zilman:2010ud,sak13}. We have previously studied the contribution of noisy cell-cycle times in driving stochastic variations of a stable protein, i.e., protein with no active degradation  \cite{sva15}, or have ignored randomness in the partitioning of molecules between the two daughters at the time of division \cite{ans14}. Exploiting the TTSHS framework,  we present a novel unified theory of how noisy cell-cycle times combine with 
randomness in the molecular partitioning process to shape variations in the level of gene product with an arbitrary decay rate. 

\subsection{Average gene product level for random cell-cycle times}
Consider a gene product synthesized at a constant rate $k_x>0$, and degrades via first-order kinetics with rate $\gamma_x>0$. Then, its level ${\boldsymbol x}(t)$ within
 the cell at time $t$ evolves as 
\begin{equation}
\frac{d\boldsymbol{x}}{dt}=k_x- \gamma_x \boldsymbol{x}(t). \label{concentration}
\end{equation}
Cell division events occur at times $\boldsymbol{t}_s, \ s\in \{1,2,\ldots\}$ with cell-cycle times $\boldsymbol \tau_s  \equiv   \boldsymbol t_s - \boldsymbol  t_{s-1}$ being iid random variables. Assuming perfect partitioning of molecules between two daughters for now, the level is exactly halved at the time of division
\begin{equation}
\boldsymbol{x}(\boldsymbol{t}_s^+)= \frac{\boldsymbol{x}(\boldsymbol{t}_s^-)}{2} \ \ {\rm with \ probability \ one}.
\end{equation}
In the context of the original TTSHS \eqref{dynamics0000}-\eqref{conditional x20} this corresponds to $A=-\gamma_x $, $\hat{a}= k_x$,
$J=1/2$ and $\hat{r}= Q= B=\hat{c}=D=0$.

Since $A=-\gamma_x<0$ and $J<1$, then as per Remark 1 the mean of ${\boldsymbol x}$ exists, and using Corollary 3.2 
\begin{equation}
\overline{ \langle \boldsymbol{x} \rangle}=   \frac{k_x}{\gamma_x } -  \frac{k_x}{2\gamma_x^2 \langle \boldsymbol \tau_s \rangle }  \frac{1 -  \langle e^{-\gamma_x  \boldsymbol \tau_s}\rangle }{1 -  \frac{1}{2}\langle e^{-\gamma_x  \boldsymbol \tau_s}\rangle }  . \label{unstable protein}
\end{equation}
If the gene product happens to be a protein whose half-life is much longer than the average cell-cycle time  ($1/\gamma_x \gg \langle \boldsymbol \tau_s \rangle $), then taking the limit $\gamma_x \to 0$ in \eqref{unstable protein} yields
\begin{equation}
\overline{ \langle \boldsymbol{x} \rangle}=  \frac{k_x\langle \boldsymbol \tau_s \rangle  \left(3+CV^2_{\boldsymbol \tau_s}\right)}{2}, \ \ \ CV^2_{\boldsymbol \tau_s}  \equiv \frac{\langle \boldsymbol \tau_s^2 \rangle - \langle \boldsymbol \tau_s \rangle^2}{\langle \boldsymbol \tau_s \rangle^2}, \label{stable protein}
\end{equation}
where $CV^2_{\boldsymbol \tau_s}$ represents the noise in cell-cycle times as quantified by its coefficient of variation squared. Note that \eqref{stable protein} could also have been derived directly from Corollary 3.3. These results exemplify the earlier point that while in general, the average gene product level depends on the entire distribution of the cell-cycle time, in some limiting cases it is completely characterized by just the first two moments of $\boldsymbol \tau_s$. Moreover, in proving Corollary 3.3 we showed that
\begin{equation} \label{tau01}
\langle \boldsymbol \tau^i \rangle = \frac{\langle \boldsymbol \tau_s^{i+1}\rangle }{(i+1)\langle \boldsymbol \tau_s\rangle}.
\end{equation}
Hence, $\langle  \boldsymbol \tau \rangle= \frac{1}{2}\langle \boldsymbol \tau_s \rangle (CV^2_{\boldsymbol \tau_s}+1)$, and the mean of a gene product in a cell in \eqref{stable protein}  can be represented as 
\begin{equation}
 \overline{\langle  {\boldsymbol{x}} \rangle} =  k_x \langle \boldsymbol \tau_s \rangle +   k_x \langle \boldsymbol \tau \rangle, \label{mean of x in tau}
\end{equation}
where the first term in the right hand side shows the products inherited from the mother cell and the latter is the products synthesized in the cell.

\begin{figure}[!tb]
	\centering
	{\includegraphics[width=1\columnwidth]{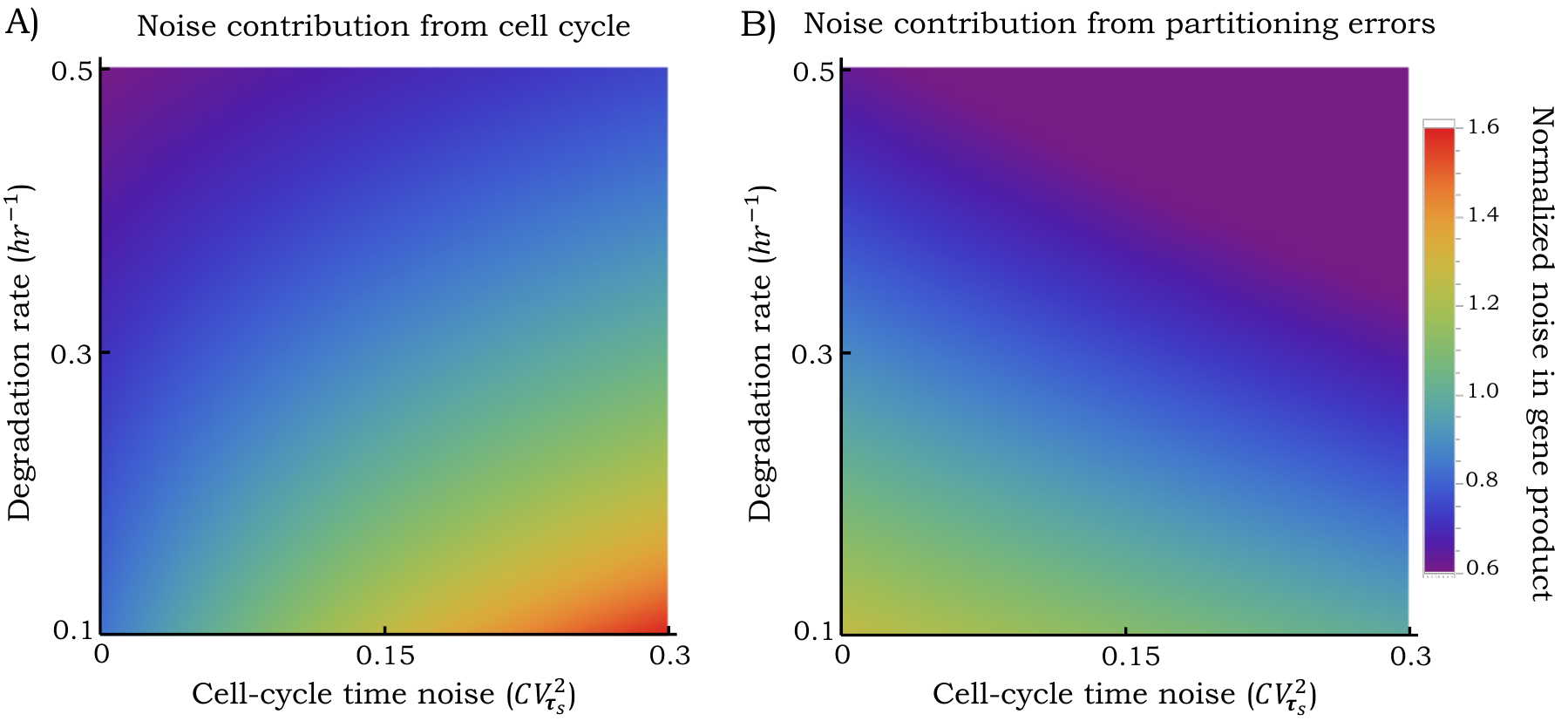}}
	\caption{{\bf The noise contributions show similar behavior with respect to decay rate, but contrasting behavior with respect to noise in cell-cycle times}. A 2D color plot
	of the two noise contributions in \eqref{totalnoise} as a function of the gene product decay rate and the noise in cell-cycle times. Increasing $CV^2_{\boldsymbol \tau_s}$ increases the noise  contribution from random cell-cycle times, but decreases the contribution form random partitioning. Both noise contributions decrease monotonically with increasing decay rate. Noise levels are normalized to their value when $CV^2_{\boldsymbol \tau_s}=0$ and $\gamma_x=0.1 \ hr^{-1}$. We used gamma distributed $\boldsymbol \tau_s$  with a fixed mean cell-cycle time of $\langle \boldsymbol \tau_s \rangle=2 hrs$. The mean of ${\boldsymbol x}$ is fixed at $100$ molecules by simultaneously changing $k_x$. }
\end{figure}

\subsection{Stochasticity in gene product levels for random cell-cycle times}
In order to calculate the second-order moments, we define a new vector $  {\boldsymbol\mu} = [ {\boldsymbol x} \ {\boldsymbol x}^2 ]^T$, whose time evolution can also be described by a TTSHS. From \eqref{mu dynamics}  it follows that
\begin{equation}
\frac{d {\boldsymbol \mu}}{dt} =\hat{a}_\mu + A_\mu   {\boldsymbol \mu}, \   A_\mu =  \left[\begin{array}{cc}
-\gamma_x  & 0\\ 2 k_x & -2\gamma_x 
\end{array}\right] ,  \ \hat{a}_\mu= \left[\begin{array}{c}
k_x\\ 0
\end{array}\right], \label{dynamics of mu}
\end{equation}
and at the time of division
 \begin{equation} \label{after division}
\langle  {\boldsymbol \mu} (\boldsymbol{t}_s^+) \rangle = J_\mu  {\boldsymbol \mu} (\boldsymbol{t}_s^-)+\hat{r}_\mu, \ \  J_\mu = \left[\begin{array}{cc}
1/2   & 0\\ 0   &   1/4
\end{array}\right] , \ \ \hat{r}_\mu  = 0 .
\end{equation}
Since $A=-\gamma_x<0$, $J=1/2$, $Q=0$, and $J\otimes J+Q\otimes Q=1/4<1$, then based on Remark 1 the steady-state second-order moment of $\boldsymbol x$ exists, and from Theorem 3.4  (see Appendix G)
\begin{equation}
	\begin{aligned}
\overline{ \langle \boldsymbol{x}^2 \rangle}=&  \frac{  k_x^2}{\gamma_x^2} +     
\frac{k_x^2 }{16 \gamma_x^3 \langle \boldsymbol \tau_s \rangle }  \frac{-14+17 \langle e^{-\gamma_x  \boldsymbol \tau_s}\rangle +  \langle e^{-2\gamma_x  \boldsymbol \tau_s}\rangle \left(2-5 \langle e^{-\gamma_x  \boldsymbol \tau_s}\rangle \right)}{\left(1-  \frac{1}{4}\langle e^{-2\gamma_x  \boldsymbol \tau_s}\rangle\right) \left(1- \frac{1}{2}\langle e^{-\gamma_x  \boldsymbol \tau_s}\rangle\right).} 
 \label{second order00}
\end{aligned}
\end{equation}
Using the coefficient of variation squared to quantify the noise in ${\boldsymbol x}$
\begin{equation}
\begin{aligned} \small
& CV^2_\text{cell cycle} \equiv \frac{\overline{ \langle \boldsymbol{x}^2 \rangle}-\overline{ \langle \boldsymbol{x} \rangle}^2}{\overline{ \langle \boldsymbol{x} \rangle}^2}\\ & = \frac{-8 \left(1 - \frac{1}{4}\langle e^{-2\gamma_x  \boldsymbol \tau_s}\rangle  \right)\left(1 -\langle e^{-\gamma_x  \boldsymbol \tau_s}\rangle  \right)^2  +4 \gamma_x \langle \boldsymbol \tau_s \rangle \left(1 - \frac{1}{4}\langle e^{-\gamma_x  \boldsymbol \tau_s}\rangle^2  \right)\left(1 -\langle e^{-2\gamma_x  \boldsymbol \tau_s}\rangle  \right)  }{8 \left(1 - \frac{1}{4}\langle e^{-2\gamma_x  \boldsymbol \tau_s}\rangle  \right)\left(- 1 +  \langle e^{-\gamma_x  \boldsymbol \tau_s}\rangle +2 \gamma_x \langle \boldsymbol \tau_s \rangle (1- \frac{1}{2}\langle e^{-\gamma_x  \boldsymbol \tau_s}\rangle)\right)^2}, \label{cv00}
\end{aligned}
\end{equation}
where $CV^2_\text{cell cycle}$ denotes the noise in the gene product level due to randomness in cell-cycle times. Before analyzing this formulas further, we next consider another physiologically relevant noise source that arises from molecular partitioning errors.

\subsection{Inclusion of randomness in the molecular partitioning process}
In reality, biomolecules in the mother cell are probabilistically partitioned between the two daughters at the time of division. For discrete number of molecules, this process is well characterized via a binomial distribution \cite{gpz05,berg_1978,rig79}. Recent work has also reported several scenarios (such as, protein multimerization) that lead to higher noise than expected from simple binomial partitioning  \cite{huh11,hup11}. Randomness in the partitioning process can be  incorporated in the  TTSHS framework with each division event resetting $ {\boldsymbol x}(\boldsymbol t_s^-)\mapsto  {\boldsymbol x}(\boldsymbol  t_s^+)$, where
\begin{align}
 \langle \boldsymbol{x}(\boldsymbol{t}_s^+)  \rangle= \frac{\boldsymbol{x} (\boldsymbol{t}_s^-)}{2},\  \
&   \langle \boldsymbol{x}^2(\boldsymbol{t}_s^+) \rangle  - \langle \boldsymbol{x}(\boldsymbol{t}_s^+) \rangle^2 =   b \boldsymbol{x} (\boldsymbol{t}_s^-).
\label{division character20}
\end{align}
Intuitively, \eqref{division character20} implies that on average, each daughter inherits half the number molecules in the mother cell, with the variance in $\boldsymbol{x}(\boldsymbol{t}_s^+) $ scaling linearly with $\boldsymbol{x} (\boldsymbol{t}_s^-)$. 
The motivation for this linear variance scaling comes from the Binomial distribution, and phenomenologically captures the notion that lower number of molecules $\boldsymbol{x} (\boldsymbol{t}_s^-)$ will lead to much higher noise in $\boldsymbol{x} (\boldsymbol{t}_s^+)$, i.e., higher coefficient of variation. The positive parameter $b$ can be interpreted as the magnitude of stochasticity in the partitioning process. 

With the above modification we have a TTSHSH where $A=-\gamma_x $, $\hat{a}= k_x$,
$J=1/2$, $B=b$, $\hat{c}=1/2$, and $\hat{r}= Q=D=0$. While the steady-state mean of gene product level is still the same as  \eqref{unstable protein}, inclusion of the nontrivial noise term in \eqref{division character20} leads to (from Theorem 3.4)
\begin{equation}
\begin{aligned}
\overline{ \langle \boldsymbol{x}^2 \rangle}=&\frac{ b k_x}{2\gamma_x^2 \langle \boldsymbol \tau_s \rangle }  \frac{1- \langle e^{-2\gamma_x  \boldsymbol \tau_s}\rangle }{1-  \frac{1}{4}\langle e^{-2\gamma_x  \boldsymbol \tau_s}\rangle }  \frac{1- \langle e^{-\gamma_x  \boldsymbol \tau_s}\rangle }{1- \frac{1}{2}\langle e^{-\gamma_x  \boldsymbol \tau_s}\rangle }    \\ & +  \frac{  k^2}{\gamma_x^2} +     
\frac{k^2 }{16 \gamma_x^3 \langle \boldsymbol \tau_s \rangle }  \frac{-14+17 \langle e^{-\gamma_x  \boldsymbol \tau_s}\rangle +  \langle e^{-2\gamma_x  \boldsymbol \tau_s}\rangle \left(2-5 \langle e^{-\gamma_x  \boldsymbol \tau_s}\rangle \right)}{\left(1-  \frac{1}{4}\langle e^{-2\gamma_x  \boldsymbol \tau_s}\rangle\right) \left(1- \frac{1}{2}\langle e^{-\gamma_x  \boldsymbol \tau_s}\rangle\right) } , \label{second order000}
\end{aligned}
\end{equation}
which yields the following elegant decomposition for gene product noise levels
\begin{equation}\label{totalnoise}
\begin{aligned} \small
&\text{Total Noise} = \frac{\overline{ \langle \boldsymbol{x}^2 \rangle}-\overline{ \langle \boldsymbol{x}^2 \rangle}}{\overline{ \langle \boldsymbol{x}^2 \rangle}} = CV^2_\text{cell cycle}+CV^2_\text{partitioning}, \\ & CV^2_\text{partitioning}  =b   \frac{1- \langle e^{-2\gamma_x  \boldsymbol \tau_s}\rangle }{1 - \frac{1}{4}\langle e^{-2\gamma_x  \boldsymbol \tau_s}\rangle }  \frac{1 -  \langle e^{-\gamma_x  \boldsymbol \tau_s}\rangle}{- 1 +  \langle e^{-\gamma_x  \boldsymbol \tau_s}\rangle +2 \gamma_x \langle \boldsymbol \tau_s \rangle (1- \frac{1}{2}\langle e^{-\gamma_x  \boldsymbol \tau_s}\rangle)}  \frac{1 }{ \overline{\langle {\boldsymbol x} \rangle} }.
\end{aligned}
\end{equation}
Here $CV^2_\text{cell cycle}$ is the noise contribution for random cell-cycle times as determined earlier, and the new term $CV^2_\text{partitioning}$, quantifies the contribution from partitioning noise. Note that unlike $CV^2_\text{cell cycle}$, $CV^2_\text{partitioning}$ is inversely related to the mean $ \overline{\langle {\boldsymbol x} \rangle}$, and would become the dominating noise term at low molecular levels. 

\begin{SCfigure}[][!tb]
	\centering
	{\includegraphics[width=0.6\columnwidth]{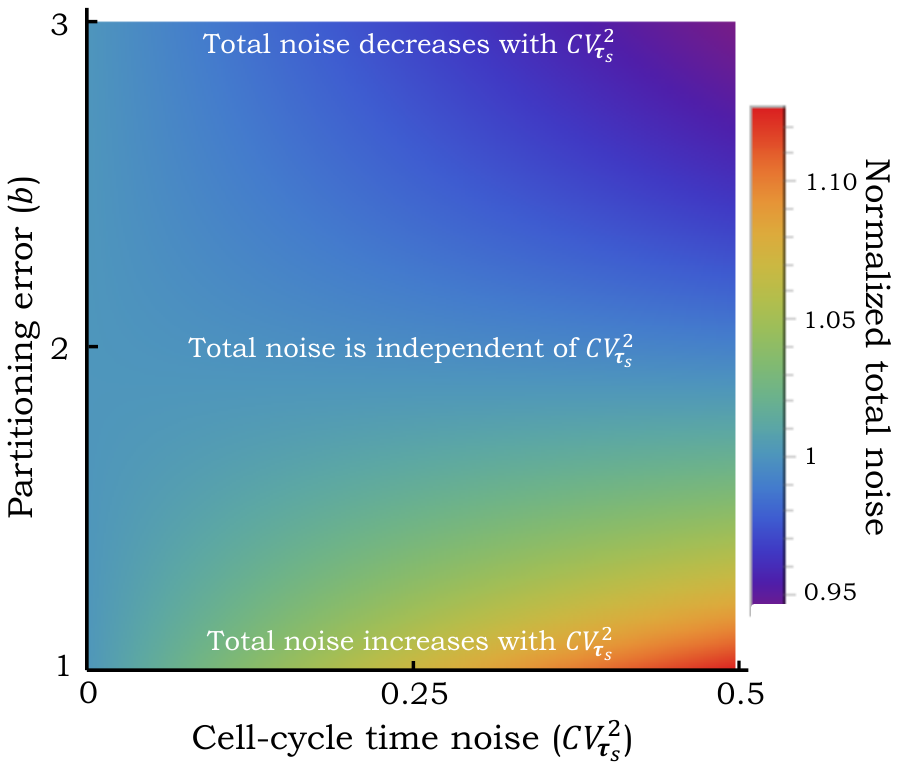}}
	\caption{{\bf Gene product noise levels can both increase or decrease with increasing noise in cell-cycle times}. The total noise in \eqref{totalnoise}
	is plotted as a function of parameter $b$ in the partitioning process and noise in cell-cycle times. While for small (large) values of $b$ the noise levels increase (decrease) with increasing $CV^2_{\boldsymbol \tau_s}$, intermediate values of $b$ can make the total noise approximately invariant of $CV^2_{\boldsymbol \tau_s}$. Noise levels are normalized to their value when $CV^2_{\boldsymbol \tau_s}=0$, the mean of ${\boldsymbol x}$ is fixed at $20$ molecules by simultaneously changing $k_x$, and $\gamma_x = 0.1 \ hr^{-1}$. The rest of parameters are chosen equal to their value in Fig. 2.}
\end{SCfigure}

Both noise contributions $CV^2_\text{cell cycle}$ and $CV^2_\text{partitioning}$  monotonically decrease to zero with increasing degradation rate $\gamma_x$, for a fixed mean $\overline{ \langle \boldsymbol{x} \rangle}$ (Fig. 2). This makes intuitive sense, as rapid turnover rates allow for faster convergence to mean levels after random perturbations.
In the limit of fast decay rate  ($\gamma_x \to \infty$), we obtain the following asymptotes
\begin{equation}
CV^2_\text{cell cycle} \approx \frac{1 }{8 \gamma_x \langle \boldsymbol \tau_s \rangle } , \ \ CV^2_\text{partitioning} \approx \frac{1}{2 \gamma_x \langle \boldsymbol \tau_s \rangle } \frac{b }{ \overline{\langle {\boldsymbol x} \rangle} }, 
\end{equation}
which only depend on the mean cell-cycle times and show very similar scaling that differ by a factor of $4b$ over mean. Interestingly, noise contributions show contrasting behavior to increasing noise in cell-cycle times -- increasing $CV^2_{\boldsymbol \tau_s}$ for fixed $ \boldsymbol \tau_s$ increases $CV^2_\text{cell cycle}$, but decreases $CV^2_\text{partitioning}$ (Fig. 2B)
This implies that depending on the degree of randomness in partitioning (parameter $b$), the total noise may decrease, increase, or remain somewhat invariant of $CV^2_{\boldsymbol \tau_s}$ (Fig. 3). Finally, taking the limit  $\gamma_x \to0$ in \eqref{totalnoise}, we recover our prior result for stable gene products  \cite{sva15}
\begin{equation} \label{stable mean}
\begin{aligned}
\text{Total noise}=\overbrace{ \frac{1}{27}+\frac{4\left(9\frac{\langle \boldsymbol \tau_s^3\rangle}{\langle \boldsymbol \tau_s  \rangle^3}-9-6CV^2_{\boldsymbol \tau_s} -7CV^4_{\boldsymbol \tau_s}\right)}{27\left(3+CV^2_{\boldsymbol \tau_s}\right)^2}}^{CV^2_\text{cell cycle}}  +\overbrace{\frac{16 b}{3(3+CV^2_{\boldsymbol \tau_s})}\frac{1}{\overline{ \langle \boldsymbol x \rangle}}}^{CV^2_\text{partitioning}},
\end{aligned}
\end{equation}
explicitly showing $CV^2_\text{partitioning}$ to be a decreasing function of  $CV^2_{\boldsymbol \tau_s}$, and the dependence of gene product noise levels on just the first three moments of $\boldsymbol \tau_s$.

\section{Linear timer-dependent TTSHS} 
While our analysis has been restricted to continuous dynamics modeled as a linear time-invariant system, we now generalize these results to time-varying systems. It is important to point out that by time varying we imply 
\begin{equation}
\frac{d {\boldsymbol x} }{dt}= \hat{a}(\boldsymbol \tau)+A(\boldsymbol \tau)  {\boldsymbol x}(t). \label{dynamics}
\end{equation}
where the vector $\hat{a}(\boldsymbol \tau)$ and the matrix $A(\boldsymbol \tau)$ vary arbitrarily with the timer state.  This extension is particularly relevant to the gene expression example discussed previously. As a newborn cell progresses through its cell cycle, it increases in size, and the number of copies of a given gene has to double before cell division. These changes in cell size and gene dosage within the cell cycle critically influence the production rates of RNAs and proteins \cite{ss16,scs15,pnb15,mmg17,zopf13,nkl15,ksc15,sxn16}, and corresponds to $k_x$ in \eqref{concentration} being timer-dependent.  Such a timer-dependent production rate is also needed for analyzing genes that are expressed at specific instants or durations within the cell cycle \cite{stk15,scs15,brb17}.
Thus, \eqref{dynamics} captures expression dynamics of a wide class of genes, and looking  beyond biology, it aids in the analysis of physical, ecological and engineering systems with time-varying dynamics. In addition to \eqref{dynamics}, we also generalize the reset value $\boldsymbol{x}(\boldsymbol{t}_s^+)$ by allowing $J(\boldsymbol \tau )$, $\hat{r}(\boldsymbol \tau )$ in \eqref{conditional x0}, and $ Q(\boldsymbol \tau )$, $ B(\boldsymbol \tau )$, $ \hat{c}(\boldsymbol \tau )$,  $D(\boldsymbol \tau )$ in \eqref{conditional x20} to be timer dependent.

\subsection{The steady-state moments of linear timer-dependent TTSHS}
Suppose that the states of the system after the $s^{th}$ reset is given by $\boldsymbol{x}(\boldsymbol t_{s}^+)$, then the states of the system for any $\boldsymbol \tau $ before the $s+1^{th}$ event are
\begin{equation}
{ {\boldsymbol{x}}}(\boldsymbol t_{s}+\boldsymbol \tau)  = \varPhi(\boldsymbol \tau ,0)\boldsymbol{x}(\boldsymbol t_{s}^+) +\int_{0}^{\boldsymbol \tau} \varPhi(\boldsymbol \tau ,l)\hat{a}(l)dl, \ \ \ \  \varPhi(\boldsymbol \tau,l)= \varPsi(\boldsymbol \tau)\varPsi^{-1}(l) \label{time-varying}
\end{equation}
\cite{tsa93}. Here $\varPsi(\boldsymbol \tau)$ is called fundamental matrix and satisfies the following
\begin{equation}
\frac{d \varPsi}{d\tau }= A(\tau)\varPsi(\tau), \ \   \vert \varPsi(\tau) \vert \neq 0,
\end{equation}
where $ \vert \  \vert$ denotes determinant of a matrix. Building upon this introduction, the following theorem gives the steady-state mean of $\boldsymbol x$.
\begin{theorem}
	The steady-state mean of the states of the timer dependent TTSHS given by \eqref{dynamics} and \eqref{reset} is 
\begin{equation}
	\begin{aligned}
	&\overline{\langle { {\boldsymbol{x}}} \rangle} = \left \langle  \varPhi(\boldsymbol \tau,0) \right  \rangle \left(I_n - \left \langle  J(\boldsymbol \tau_s) \varPhi(\boldsymbol \tau_s,0)  \right  \rangle  \right)^{-1} \times \\
	& \hspace{10mm} \left(\left \langle J(\boldsymbol \tau_s) \int_{0}^{\boldsymbol \tau_s}\varPhi(\boldsymbol \tau_s,l)\hat{a}(l)dl \right \rangle +\langle  \hat{r}(\boldsymbol \tau_s) \rangle  \right)  +\left \langle  \int_{0}^{\boldsymbol \tau}\varPhi(\boldsymbol \tau,l)\hat{a}(l)dl \right \rangle
	\end{aligned}
	\label{mean of xx}	
	\end{equation}	
if and only if $
\left\langle \varPhi(\boldsymbol \tau_s,0)  \right \rangle $ exist and all the eigenvalues of $ \left\langle J(\boldsymbol \tau_s)\varPhi(\boldsymbol \tau_s,0) \right \rangle$ are inside the unit circle. 
\end{theorem}
Please see Appendix H for the proof. Here we use the notation $\boldsymbol \tau_s$ when we take expected value with respect to $\boldsymbol \tau_s$ (e.g. $\left\langle \varPhi(\boldsymbol \tau_s,0)  \right \rangle $) and we use $\boldsymbol \tau$ when we take expected value with respect to $\boldsymbol \tau$ (e.g. $\left\langle \varPhi(\boldsymbol \tau,0)  \right \rangle $). 
Note that for a general $A(\boldsymbol \tau)$, one needs to calculate $\varPhi$ for obtaining mean of $ {\boldsymbol x}$. However, except few cases, closed form of $\varPhi$ does not exist \cite{tsa93}. One of these few cases is $A(\boldsymbol \tau)=0$, where $\varPhi$ in \eqref{mean of xx} is simply $I_n$. In this case 
\begin{equation}
\begin{aligned}
& \overline{\langle  {\boldsymbol{x}} \rangle}=   \left(I_n -\left\langle  J(\boldsymbol \tau_s) \right \rangle  \right) ^{-1}
\left( \left \langle J(\boldsymbol \tau_s)   \int_0^{\boldsymbol \tau_s} \hat{a}(l) dl  \right \rangle   + \langle \hat{r}(\boldsymbol \tau_s) \rangle \right)   + \left \langle  \int_0^{\boldsymbol \tau}  \hat{a}(l) dl\right \rangle . 
\end{aligned}
\end{equation} 	
This equation simplifies to \eqref{mean of x A=0} for time-invariant $J$, $\hat{r}$ and $\hat{a}$. Another limit in which the matrix $\varPhi$ can be derived easily is when $A( \tau)$ and $e^{\int_0^{\tau } A(y) d y}$ can commute, in this case 
\begin{equation} \label{varphi}
\varPhi(\boldsymbol \tau,0)= e^{\int_{0}^{\boldsymbol  \tau} A(y) d y} , \ \ \int_{0}^{\boldsymbol  \tau}\varPhi(\boldsymbol  \tau,l)\hat{a}(l)dl=e^{\int_0^{\boldsymbol  \tau} A(y) d y}  \int_0^{\boldsymbol \tau}  e^{-\int_0^{l} A(y) d y}   \hat{a}(l) dl ,
\end{equation}
and as a result \eqref{mean of xx} simplifies to
\begin{equation}
\begin{aligned}
\overline{\langle { {\boldsymbol{x}}} \rangle} = &\left \langle   e^{\int_{0}^{\boldsymbol \tau} A(y) d y} \right  \rangle \left(I_n - \left \langle  J(\boldsymbol \tau_s)  e^{\int_{0}^{\boldsymbol \tau_s} A(y) d y}  \right  \rangle  \right)^{-1} \times \\
&  \left(\left \langle J(\boldsymbol \tau_s)  e^{\int_0^{\boldsymbol \tau_s} A(y) d y}  \int_0^{\boldsymbol \tau_s}  e^{-\int_0^{l} A(y) d y}   \hat{a}(l) dl   \right \rangle +\langle  \hat{r}(\boldsymbol \tau_s) \rangle  \right)  \\ & +\left \langle  e^{\int_0^{\boldsymbol \tau} A(y) d y}  \int_0^{\boldsymbol \tau}  e^{-\int_0^{l} A(y) d y}   \hat{a}(l) dl   \right \rangle.
\end{aligned}	
\end{equation}
Examples of this case include $A(\boldsymbol \tau)$ being diagonal (dynamics of each state depends only on itself), and if $A(\boldsymbol \tau)=A k(\boldsymbol \tau)$, where $A$ is a constant matrix and $k(\boldsymbol \tau)$ is a scalar time-varying function.

Moreover, similar to Section 3.2, we can define the vector 
$\small {\boldsymbol \mu } \equiv \left[ {\boldsymbol x}^\top \ \ \ {\rm vec}  \left(  {\boldsymbol x}   {\boldsymbol x}^\top \right)^\top\right]^\top$, where its dynamics between the events is given by
\begin{align}
&\frac{d {\boldsymbol \mu }}{dt} =\hat{a}_\mu (\boldsymbol \tau )+ A_\mu (\boldsymbol \tau )   {\boldsymbol \mu }.
\end{align}
Here $\hat{a}_\mu (\boldsymbol \tau )$ and $A_\mu (\boldsymbol \tau )$ are similar to \eqref{Auau0} for time-varying $\hat{a} (\boldsymbol \tau )$ and $A (\boldsymbol \tau )$. 
This system is in the form of \eqref{dynamics} hence its solution between the events is similar to \eqref{time-varying} for appropriate $\varPhi_\mu$. Further, during an event, the states of vector $ {\boldsymbol\mu}$ change as \eqref{reset 2} where mean of $ {\boldsymbol \mu}(\boldsymbol{t}_s^+)$ is related to $ {\boldsymbol \mu}(\boldsymbol{t}_s^-)$ as
\begin{align}
&\hspace{25mm}\langle  {\boldsymbol \mu }(\boldsymbol{t}_s^+)  \rangle =   J_\mu (\boldsymbol \tau )   {\boldsymbol \mu }(\boldsymbol{t}_s^-) + \hat{r}_\mu (\boldsymbol \tau ) .
\end{align}
In this equation $J_\mu (\boldsymbol \tau )$ and $\hat{r}_\mu (\boldsymbol \tau )$ are time-varying counterpart of $J_\mu$ and $\hat{r}_\mu$ in \eqref{Jmu}. 

Given this reformulation, similar to Theorem 3.4, we can derive the second-order moments of $\boldsymbol x$ through the vector $\boldsymbol \mu$. The steady-state mean of $\boldsymbol \mu$ is
	\begin{equation} 
		\begin{aligned}
		&\overline{\langle  {\boldsymbol \mu }  \rangle}=  \left \langle  \varPhi_\mu(\boldsymbol \tau,0) \right  \rangle \left(I_n - \left \langle  J_\mu(\boldsymbol \tau_s) \varPhi_\mu (\boldsymbol \tau_s,0)  \right  \rangle  \right)^{-1} \times \\
		& \hspace{5mm} \left(\left \langle J_\mu(\boldsymbol \tau_s) \int_{0}^{\boldsymbol \tau_s}\varPhi_\mu (\boldsymbol \tau_s,l)\hat{a}_\mu(l)dl \right \rangle +\langle  \hat{r}_\mu(\boldsymbol \tau_s) \rangle  \right)  +\left \langle  \int_{0}^{\boldsymbol \tau}\varPhi_\mu(\boldsymbol \tau,l)\hat{a}_\mu(l)dl \right \rangle
		\end{aligned} \label{mean mu} 
	\end{equation}
if and only if all the eigenvalues of the matrix $ \langle ( J(\boldsymbol \tau_s) \otimes  J(\boldsymbol \tau_s)+Q(\boldsymbol \tau_s)\otimes  Q(\boldsymbol \tau_s) )$ 
$ (  \varPhi(\boldsymbol \tau_s,0)  \otimes  \varPhi(\boldsymbol \tau_s,0)) \rangle $ are inside the unit circle. Furthermore, this is straightforward to see that \eqref{varphi} can be extended to vector $\boldsymbol \mu$, i.e., if $A_\mu(\tau )$ and $ e^{\int_0^{\tau } A_\mu(y) d y}$ can commute then 
	\begin{equation} \label{extension of 53}
\varPhi_\mu(\boldsymbol  \tau,0)= e^{\int_{0}^{\boldsymbol  \tau} A_\mu(y) d y} ,\ \int_{0}^{\boldsymbol  \tau}\varPhi_\mu(\boldsymbol  \tau,l)\hat{a}_\mu(l)dl=e^{\int_0^{\boldsymbol  \tau} A_\mu(y) d y}  \int_0^{\boldsymbol \tau}  e^{-\int_0^{l} A_\mu(y) d y}   \hat{a}_\mu(l) dl   .
\end{equation}
Finally, Remark 1 also can be generalized to timer-dependent case if 1- $A(\boldsymbol \tau)$ is a negative definite matrix for all the values of $\boldsymbol \tau$, 2- The matrices $J(\boldsymbol \tau)$ and $ J(\boldsymbol \tau)\otimes  J(\boldsymbol \tau)+Q(\boldsymbol \tau)\otimes Q(\boldsymbol \tau)$ are diagonal positive definite and all of their eigenvalues are inside the unit circle. Then the first and second-order moments of $\boldsymbol x$ exists irrespective of distribution of $\boldsymbol \tau_s $. In the next part, we use our results to study time-varying synthesis rate.
\subsection{Timer-dependent gene expression dynamics}
Revisiting the gene expression example, \eqref{concentration} is now modified as 
\begin{equation}
\frac{d\boldsymbol{x}}{dt}=k_x (\boldsymbol \tau)- \gamma_x \boldsymbol{x}(t),
\end{equation}
where $k_x (\boldsymbol \tau)$ represents a generalized timer-dependent production rate. Assuming the same structure of resets as in \eqref{division character20}, Theorem 5.1 yiels
\begin{equation} \label{54}
\begin{aligned}
& \overline{\langle  {\boldsymbol{x}} \rangle}= \frac{\left \langle e^{-\gamma_x \boldsymbol \tau_s } \right \rangle}{\gamma_x \langle \boldsymbol \tau_s \rangle }\frac{1-\left \langle e^{-\gamma_x \boldsymbol \tau_s } \right \rangle}{2-\left \langle e^{-\gamma_x \boldsymbol \tau_s } \right \rangle}
\left \langle \int_0^{\boldsymbol \tau_s}   e^{\gamma_x l } k_x(l) dl\right \rangle  + \left \langle e^{-\gamma_x \boldsymbol \tau } \int_0^{\boldsymbol \tau}  e^{\gamma_x l }  k_x(l) dl\right \rangle .
\end{aligned}
\end{equation}
In the limit of constant synthesis rate $k_x(\boldsymbol \tau)=k_x $ 
\begin{subequations}
\begin{align}
\left \langle \int_0^{\boldsymbol \tau_s}   e^{\gamma_x l } k_x(l) dl\right \rangle  = k_x \left \langle \int_0^{\boldsymbol \tau_s}   e^{\gamma_x l }dl\right \rangle = \frac{k_x}{\gamma_x} \left \langle e^{\gamma_x \boldsymbol \tau_s } \right \rangle,\\
\left \langle e^{-\gamma_x \boldsymbol \tau } \int_0^{\boldsymbol \tau}  e^{\gamma_x l }  k_x(l) dl\right \rangle = k_x \left \langle e^{-\gamma_x \boldsymbol \tau } \int_0^{\boldsymbol \tau}  e^{\gamma_x l }  dl\right \rangle= \frac{k_x}{\gamma_x} .
\end{align} \label{55}
\end{subequations}
By putting \eqref{55} back in \eqref{54}, the mean of $\boldsymbol x$ simplifies to \eqref{53} for constant synthesis rate. Moreover,  by taking the limit  $\gamma_x \to0$ in \eqref{54}, we obtain the mean of stable gene products
\begin{equation}
\begin{aligned}
& \overline{\langle  {\boldsymbol{x}} \rangle}= 
\left \langle  \int_0^{\boldsymbol \tau_s} k_x(l) dl  \right \rangle   + \left \langle  \int_0^{\boldsymbol \tau}  k_x(l) dl\right \rangle . 
\end{aligned}\label{varying mean0}
\end{equation}
Interestingly, from \eqref{varying mean0} it follows that for a given constant mean level of $\boldsymbol x$, high values of $k_x(\boldsymbol \tau)$ when $\boldsymbol \tau$ is small (beginning of cell cycle) results in lower $\langle k_x(\boldsymbol \tau) \rangle $. This means that production at the beginning of cell cycle needs less production events and less resources to keep a given mean of $\boldsymbol x$. Finally, suppose that $k_x(\boldsymbol \tau)=k_x $ then
\begin{equation}
\begin{aligned}
&  \overline{\langle  {\boldsymbol{x}} \rangle}= 
\left \langle  \int_0^{\boldsymbol \tau_s} k_x(l) dl  \right \rangle   + \left \langle  \int_0^{\boldsymbol \tau}  k_x(l) dl\right \rangle =  k_x (\langle \boldsymbol \tau_s \rangle +  \langle \boldsymbol \tau \rangle),
\end{aligned} \label{53}
\end{equation}
which is equal to \eqref{mean of x in tau}.

For a general $k_x(\boldsymbol \tau)$ providing the analytic formulas for noise of an unstable gene product is convoluted. On the other hand, for a stable product ($\gamma_x\approx 0$) the noise contribution from cell-cycle time variations and partitioning errors is
\begin{subequations}
\label{varying k}
\begin{align}
&CV^2_\text{cell cycle}=  \frac{ 2 \left \langle \int_0^{\boldsymbol \tau_s} \left( k_x(\tau)  \int_0^{\tau }  k_x(l) dl \right)d\tau \right \rangle -\left \langle  \int_0^{\boldsymbol \tau_s}  k_x(l) dl\right \rangle^2  }{3 \left(\left \langle  \int_0^{\boldsymbol \tau_s} k_x(l) dl  \right \rangle   + \left \langle  \int_0^{\boldsymbol \tau}  k_x(l) dl\right \rangle\right)^2} \\  \nonumber & \hspace{20 mm}+
 \frac{  2 \left \langle \int_0^{\boldsymbol \tau } \left( k_x(\tau)  \int_0^{\tau }  k_x(l) dl \right)d\tau \right \rangle- \left \langle  \int_0^{\boldsymbol \tau}  k_x(l) dl\right \rangle^2  }{ \left(\left \langle  \int_0^{\boldsymbol \tau_s} k_x(l) dl  \right \rangle   + \left \langle  \int_0^{\boldsymbol \tau}  k_x(l) dl\right \rangle\right)^2},\\
& CV^2_\text{partitioning}= \frac{8 b }{3}\frac{\left \langle  \int_0^{\boldsymbol \tau_s} k_x(l) dl  \right \rangle }{\left \langle  \int_0^{\boldsymbol \tau_s} k_x(l) dl  \right \rangle   + \left \langle  \int_0^{\boldsymbol \tau}  k_x(l) dl\right \rangle}\frac{1}{\overline{ \langle \boldsymbol x \rangle}}.
\end{align}
\end{subequations}
These results simplify to \eqref{stable mean} for a constant synthesis rate. 

As an example, we study protein count and noise in a mammalian cell. The volume of mammalian cells is highly variable within a population \cite{tkl09,bga10}. However, a key necessity for maintaining cellular functions is to keep concentartion of different proteins constant. This means that the number of proteins should scale with the volume of cell \cite{mb12,zlw10,msc12}. To cover this case, consistent with measurements, we assume that the synthesis rate scales with the cell volume which is an exponential function of the cell cycle \cite{vgs18}. Thus, we assume that synthesis rate $k_x (\boldsymbol \tau)$ is exponentially increasing throughout the cell cycle and eventually doubles at the end of cell cycle  \cite{wnt16}; the synthesis rate is 
\begin{equation}
k_x (\boldsymbol \tau) = k_x e^{\frac{ln(2)\boldsymbol \tau}{\langle \boldsymbol \tau_s \rangle }}, \label{k exp}
\end{equation}
where $k_x$ is a non-negative constant. Since a large number of proteins in mammalian cells are stable \cite{sbl11}, we do not need to consider the degradation of $\boldsymbol x$. Hence we can replace \eqref{k exp} in \eqref{varying mean0} to derive mean
\begin{equation}
\overline{\langle  {\boldsymbol{x}} \rangle}= \frac{ k_x  \langle \boldsymbol \tau_s \rangle  \left((1+\ln(2)) \left \langle 2^{ \frac{\boldsymbol \tau_s}{\langle \boldsymbol \tau_s \rangle}} \right \rangle -1-2\ln (2)\right)}{\ln^2(2)}, \ \ \left \langle 2^{ \frac{\boldsymbol \tau_s}{\langle \boldsymbol \tau_s \rangle} }\right \rangle = \int_0^{\infty}f(\tau) 2^{\frac{ \tau}{ \langle \boldsymbol \tau_s \rangle }} d \tau .
\end{equation}
While for constant synthesis rate, mean of a stable gene product just depend on the mean and noise of cell-cycle times, for exponentially increasing production rate the mean of $\boldsymbol x$ depends on the entire distribution of cell-cycle times.

\begin{SCfigure}[][!t]
	\centering
	{\includegraphics[width=0.6\columnwidth]{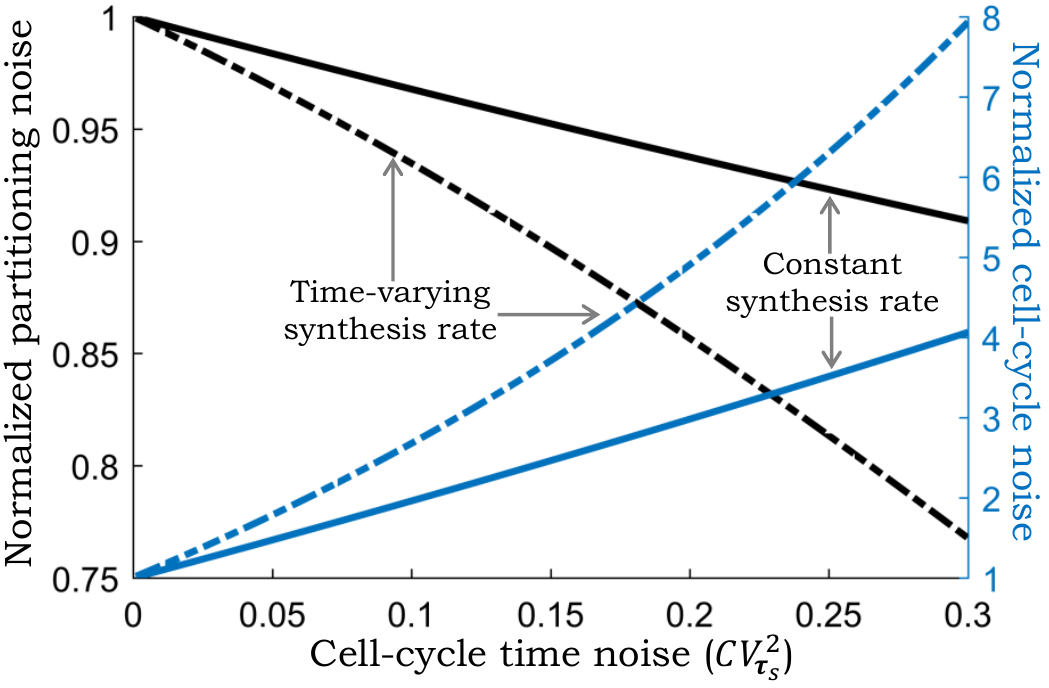}}
	\caption{{\bf The noise in a gene product is more sensitive to cell-cycle time variations when the synthesis rate is not constant}. For a time-varying synthesis rate, the noise contribution from partitioning errors and cell-cycle time are affected more by $CV^2_{\boldsymbol \tau_s}$ when $k_x(\boldsymbol \tau)$ is timer-dependent. Noise levels are normalized to their value when $CV^2_{\boldsymbol \tau_s}=0$. The mean cell-cycle time is $\boldsymbol \tau_s=20 \ mins$ (fast growing bacteria) and the rest of parameters are chosen equal to their value in Fig. 2.}
\end{SCfigure}

In the next step, we calculate the noise contribution from cell-cycle time variations and partitioning errors by replacing \eqref{k exp} in \eqref{varying k}
\begin{subequations}\label{noise varying case}
	\begin{align}
	&	CV^2_\text{cell cycle}=\frac{\ln(2) \left( \left \langle 4^{ \frac{\boldsymbol \tau_s}{\langle \boldsymbol \tau_s \rangle}}\right \rangle  -1\right)-2 \left( \left \langle2^{ \frac{\boldsymbol \tau_s}{\langle \boldsymbol \tau_s \rangle}}\right \rangle  -1\right)^2}{2 \left({  (1+\ln(2))\left  \langle 2^{ \frac{\boldsymbol \tau_s}{\langle \boldsymbol \tau_s \rangle}} \right  \rangle -1-2\ln (2)}\right)^2}, \label{noise varying case01}\\
	&	CV^2_\text{partitioning}=\frac{8b \ln(2) \left( \left \langle 2^{ \frac{\boldsymbol \tau_s}{\langle \boldsymbol \tau_s \rangle}} \right \rangle  -1 \right)   }{3\left(  (1+\ln(2)) \left \langle 2^{ \frac{\boldsymbol \tau_s}{\langle \boldsymbol \tau_s \rangle}} \right\rangle -1-2\ln (2)\right) } \frac{1}{\overline{ \langle \boldsymbol x \rangle}}.\label{noise varying case02}
	\end{align}
\end{subequations}
Our analysis shows that these noise terms are more affected by statistical characteristics of cell-cycle time than the case of constant synthesis rate (Fig. 4). This means that keeping concentration constant will make cells more vulnerable to cell-cycle noise. However, in the limit of large $b$, a cell can exploit this dependency to reduce the contribution of noisy cell-cycle times (Fig. 4). Moreover, in the limit of deterministic cell-cycle times, noise values in \eqref{noise varying case} are slightly higher than those of constant synthesis rate in \eqref{stable mean}. This implies that keeping the concentration constant also may come with the price of having higher noise levels in the count level. 

In addition to mammalian cells, measurements shows that in fast growing bacteria the synthesis rate is continuously increasing \cite{wnt16}. In these cells multiple gene replication occurs
throughout the cell cycle and the amount of other components required for expression (e.g. RNA polymerases and ribosomes) is limited. Hence, maybe considering a linearly increasing instead of exponentially increasing synthesis rate is more physiological. Our analysis reveals that in this case the noise behavior is similar to Fig. 4 qualitatively.

Finally, our analytical results provide a unique method for inferring partitioning noise in a gene product. Current expreiments are able to quantify distribution of cell-cycle time \cite{wrp10}. By Measuring noise in a gene product we can use \eqref{noise varying case01} to calculate the contribution of cell-cycle time variations on total noise. The subtraction of total noise from \eqref{noise varying case01} provides noise contribution from partitioning and hence it can be used to infer the partitioning scenario of a specific protein (parameter $b$).


\section{Conclusion}
Moment analysis of Stochastic Hybrid Systems (SHS) often relies on deriving a set of differential equations for the time evolution of moments  \cite{hsi04,sih10a}. For linear stochastic systems, moments can be obtained exactly by solving these set of differential equations. However,
nonlinearities within SHS, such as the hazard rate \eqref{hr}, lead to unclosed dynamics in the sense that time evolution of lower-order moments depends on higher-order moments. In such cases, moment computations are performed by either employing approximate closure schemes \cite{whi57,kcm05,kon12,smk13,gri12,sih10, svs15,sih05,gsl17,ddd16}, or constraints imposed by positive semidefiniteness of moment matrices \cite{gvl17,lamperski2017analysis,gls18}. 

Instead of relying on moment dynamics, here we used an alternative approach to derive exact analytical expressions for the first two steady-state moments of TTSHS. Our main results (Theorems 3.1, 3.4, and 5.1) connect these moments to the system dynamics and the distribution of event arrival times. While knowledge of the entire distribution of 
$\boldsymbol \tau_s$ is generally needed to compute the moments, but if $A=0$ then the mean of ${\boldsymbol x}$ just depends on the first two moments of $\boldsymbol \tau_s$, and the second-order moments of ${\boldsymbol x}$ depend on the first three moments of $\boldsymbol \tau_s$ (Corollary 3.3 and Appendix F). Interestingly,  if $A$ is Hurwitz, and the resets only add 
a zero-mean noise term that can be state-dependent,  then the extent of random fluctuations in $ {\boldsymbol x}$ is only affected by the average frequency of events $1/\langle \boldsymbol \tau_s \rangle $ (equation \eqref{noisy reset}). Analogous results were derived for time-varying TTSHS where $\hat{a}$ and $A$ vary with the timer in between events. Finally, applying the theory of TTSHS to the biological example of gene expression resulted in novel formulas for the mean and variance in the level of a gene product, and how these levels are impacted by stochasticity in cell-cycle times and the molecular partitioning process. 

Future works will extend our method to consider TTSHS where continuous dynamics follow a stochastic differential equation, or multi-mode TTSHS that allow for stochastic switching
between linear systems. Recent work has shown that for some nonlinear stochastic systems moment dynamics become automatically closed at some higher-order moments, and hence moments can be computed exactly in spite of unclosed moment dynamics of lower-order moments \cite{sos15}. It will be interesting to explore classes of TTSHS with nonlinear continuous dynamics, or state-dependent event arrival rates for which moments can be computed exactly. 

\appendix
\section{Proof of Theorem 3.1} 
Using \eqref{dynamics0000}, the states of TTSHS right before $s^{th}$ event ${ {\boldsymbol{x}}}(\boldsymbol t_{s}^-)$ is related to the states of TTSHS right after $s-1^{th}$ event ${ {\boldsymbol{x}}}(\boldsymbol t_{s-1}^+) $ as  
\begin{equation}\begin{aligned}
 { {\boldsymbol{x}}}(\boldsymbol t_{s}^-)  =  e^{ A \boldsymbol \tau_s} \int_0^{\boldsymbol \tau_s}  e^{ -Al} \hat{a} dl 
+  e^{ A \boldsymbol \tau_s } { {\boldsymbol{x}}}(\boldsymbol t_{s-1}^+)     .  \label{right before}
\end{aligned}
\end{equation}
Thus, by using \eqref{dynamics0000}, the mean of the states after $s^{th}$ event is  
\begin{equation}\begin{aligned}
\langle { {\boldsymbol{x}}}(\boldsymbol t_{s}^+)   \rangle = J \left \langle e^{ A \boldsymbol \tau_s} \int_0^{\boldsymbol \tau_s}  e^{ -Al} \hat{a} dl \right \rangle 
+ J \left \langle e^{ A \boldsymbol \tau_s } \right \rangle  \left \langle  { {\boldsymbol{x}}}(\boldsymbol t_{s-1}^+)   \right \rangle  + \hat{r}. 
\end{aligned}\label{xi0}
\end{equation} 
In order to have a finite $\langle { {\boldsymbol{x}}}(\boldsymbol t_{s}^+)   \rangle$ in \eqref{xi0}, $\left \langle e^{ A \boldsymbol \tau_s} \int_0^{\boldsymbol \tau_s}  e^{ -Al} \hat{a} dl \right \rangle $ and $\left \langle e^{ A \boldsymbol \tau_s } \right \rangle  $ should be finite. In the next, we show that $\left \langle e^{ A \boldsymbol \tau_s } \right \rangle $ being finite means that $\left \langle e^{ A \boldsymbol \tau_s} \int_0^{\boldsymbol \tau_s}  e^{ -Al} \hat{a} dl \right \rangle $ is also finite. 

The fact that a matrix exponential $e^{ A \tau} $ can be written as 
\begin{equation} \label{matrix exponential}
  e^{ A \tau }   = \sum_{i=0}^{\infty}A^i \frac{ \tau ^i  }{i!}
\end{equation}
means that $A$ and $e^{ A \tau } $ can commute. Thus
\begin{equation}\label{integral2}
\begin{aligned}
A \left \langle  e^{ A\boldsymbol \tau_s} \int_0^{\boldsymbol \tau_s}  e^{ -Al} \hat{a} dl  \right \rangle & = A \int_0^{\infty}f(\tau)   e^{ A \tau }  
\int_0^\tau  e^{- As} \hat{a} dl  d \tau \\  & 
=  \int_0^{\infty} f(\tau)    e^{ A \tau }  
\int_0^\tau  e^{- As} A \hat{a} dl  d \tau \\  &
 = \int_0^{\infty}   f(\tau) e^{ A \tau }  (I_n -e^{ - A \tau }   )\hat{a}  d \tau   \\ &
 =-( I_n - \left \langle  e^{ A\boldsymbol \tau_s}  \right \rangle ) \hat{a},
\end{aligned} 
\end{equation} 
and existence of $\left \langle e^{ A \boldsymbol \tau_s } \right \rangle $ means that all the terms in \eqref{xi0} are finite so $\langle { {\boldsymbol{x}}}(\boldsymbol t_{s}^+)   \rangle$ is finite if and only if $\left \langle e^{ A \boldsymbol \tau_s } \right \rangle $ is finite. 

Moreover, from \eqref{xi0} the mean of the states right after an event in steady-state ($s\rightarrow \infty$) exists if and only and if eigenvalues of $J \left\langle  e^{ A\boldsymbol \tau_s} \right \rangle$ are inside the unite circle. In this limit the steady-state mean of the states ($s\rightarrow \infty$) right after an event can be written as
\begin{equation}
\begin{aligned}
\lim_{s\to\infty}\langle { {\boldsymbol{x}}}(\boldsymbol t_{s}^+)   \rangle= &  \left(I_n -J \left\langle e^{ A\boldsymbol \tau_s} \right \rangle   \right)^{-1} \left( J\left \langle  e^{ A\boldsymbol \tau_s} \int_0^{\boldsymbol \tau_s}  e^{ -Al} \hat{a} dl  \right \rangle  + \hat{r} \right)  .
\end{aligned}
\label{mean of xtau0}	
\end{equation}

By using equation \eqref{dynamics0000} and \eqref{mean of xtau0}, the steady-state mean of the states in between events for any values of $\boldsymbol \tau$ is
\begin{equation}
\begin{aligned}
 \lim_{s \rightarrow \infty}\langle { \boldsymbol{x}}(\boldsymbol t_{s}+\boldsymbol \tau)  \rangle= & e^{A\boldsymbol \tau } \left(I_n -J \left\langle e^{ A\boldsymbol \tau_s} \right \rangle   \right)^{-1}\left( J\left \langle  e^{ A\boldsymbol \tau_s} \int_0^{\boldsymbol \tau_s}  e^{ -Al} \hat{a} dl  \right \rangle  +\hat{r} \right) \\  & +  e^{ A\boldsymbol \tau  } \int_0^{\boldsymbol\tau }  e^{ -Al} \hat{a} dl  .
\end{aligned}
\label{first x}
\end{equation}
The mean of the states can be obtained by taking the expected value of $\langle { \boldsymbol{x}}(\boldsymbol t_{s}+\boldsymbol \tau)  \rangle$ in \eqref{first x} with respect to the all values of $\boldsymbol \tau $ by using \eqref{prob. tau}. However, to have a finite $ \overline{\langle  {\boldsymbol x}\rangle }$ we need to show that $ \left\langle  e^{ A\boldsymbol \tau} \right \rangle$ and $\left \langle  e^{ A\boldsymbol \tau} \int_0^{\boldsymbol \tau }  e^{ -Al} \hat{a} dl  \right \rangle$ are also finite. 

We show that when all the elements of $ \left\langle  e^{ A\boldsymbol \tau_s} \right \rangle$ are bounded then $\left\langle e^{ A\boldsymbol \tau} \right \rangle $ exists and is finite
\begin{equation} 
\begin{aligned}
\label{cond. on AT}
& \left\langle  e^{ A\boldsymbol \tau_s} \right \rangle = \int_0^{\infty} h(\tau) e^{-\int_0^\tau h(y) d y}  e^{A\tau }d \tau \\ & =  \left(  -e^{-\int_0^{\tau} h(y) d y} e^{A\tau }   \right)_0^\infty + \int_0^{\infty}    e^{-\int_0^{\tau} h(y) d y}e^{A\tau }A  d\tau =  I_n +\langle \boldsymbol \tau_s \rangle \left\langle e^{ A\boldsymbol \tau} \right \rangle A,
\end{aligned}
\end{equation}
where we used the fact that $\lim_{\tau\rightarrow \infty} e^{-\int_0^{\tau} h(y) d y} e^{A\tau }=0 $. For the sake of simplicity of mathematical notation we proof this for scalar case of $A=a$. From \eqref{prob. tau} it follows that 
\begin{equation}
\int_0^\infty p(\tau) d\tau=1 \Rightarrow \int_0^\infty  e^{-\int_0^{\tau} h(y) d y} d\tau  = \langle  \boldsymbol \tau_s \rangle < \infty \Rightarrow \lim_{\tau\rightarrow \infty}  e^{-\int_0^{\tau} h(y) d y}= 0. 
\end{equation}
In the next, assume that $\lim_{\tau\rightarrow \infty}  e^{ a \tau} $ is infinite, hence
\begin{equation}
 \lim_{\tau\rightarrow \infty}  e^{-\int_0^{\tau} h(y) d y}  e^{ a \tau}  = 0 \times \infty.  
\end{equation}
We use l'Hopital's rule 
\begin{equation}
\lim_{\tau\rightarrow \infty}  e^{-\int_0^{\tau} h(y) d y}  e^{ a \tau}  = - \frac{1}{a} \lim_{\tau\rightarrow \infty} h(\tau)e^{-\int_0^{\tau} h(y) d y}  e^{ a \tau}  .  
\end{equation}
Finally, note that we assumed moment generating function exists, hence 
\begin{equation}
\left\langle  e^{ a\boldsymbol \tau_s} \right \rangle < \infty \Rightarrow \lim_{\tau\rightarrow \infty} h(\tau)  e^{-\int_0^{\tau} h(y) d y} e^{a\tau }= 0
\end{equation}
and this completes our proof. 

Moreover, similar to \eqref{integral2} we have
	\begin{align}
	& A \left \langle  e^{ A\boldsymbol \tau} \int_0^{\boldsymbol \tau }  e^{ -Al} \hat{a} dl  \right \rangle =-( I_n - \left \langle  e^{ A\boldsymbol\tau }  \right \rangle ) \hat{a}. 
	\end{align}
Hence, existence of $ \left\langle  e^{ A\boldsymbol \tau_s} \right \rangle$ means that all the matrices in \eqref{mean of x} exists. 

\section{Proof of Corollary 3.2}  
Taking integral by parts, $\left\langle e^{ A\boldsymbol \tau} \right \rangle $ can be written as
\begin{equation}
\begin{aligned}
\left\langle e^{ A\boldsymbol \tau} \right \rangle = \frac{1}{\langle \boldsymbol \tau_s \rangle }\int_0^{\infty}   e^{-\int_0^{\tau} h(y) d y} e^{A\tau }d\tau =  \frac{1}{\langle \boldsymbol \tau_s \rangle }\left(  e^{-\int_0^{\tau} h(y) d y} e^{A\tau } A^{-1}   \right)_0^\infty \\ +  \frac{1}{\langle \boldsymbol \tau_s \rangle }\int_0^{\infty}  h(\tau)  e^{-\int_0^{\tau} h(y) d y}e^{A\tau }A^{-1}   d\tau =  \frac{-1}{\langle \boldsymbol \tau_s \rangle }\left(  I_n - \left\langle e^{ A  \boldsymbol  \tau_s}  \right \rangle \right)A^{-1}.
\end{aligned}
\end{equation}
Moreover 
\begin{equation}
\left \langle  e^{ A\boldsymbol \tau_s} \int_0^{\boldsymbol \tau_s}  e^{ -Al} \hat{a} dl  \right \rangle = \left \langle  e^{ A\boldsymbol \tau_s} (I_n-e^{ - A \boldsymbol \tau_s }   )A^{-1} \hat{a}    \right \rangle  =-( I_n - \left \langle  e^{ A\boldsymbol \tau_s}  \right \rangle ) A^{-1} \hat{a}  . 
\end{equation}
Finally, the last integral in \eqref{mean of x} can be written as
\begin{equation}
\begin{aligned}
\int_0^{\infty}  e^{-\int_0^{\tau} h(y) d y}   e^{ A \tau }  
\int_0^\tau  e^{- As} \hat{a} dl  d \tau &  = \int_0^{\infty}  e^{-\int_0^{\tau} h(y) d y}   e^{ A \tau }  (I_n -e^{ - A \tau }   )A^{-1} \hat{a}  d \tau   \\ & = -\left(I_n- \left\langle e^{ A  \boldsymbol  \tau_s}  \right \rangle \right) A^{-2} \hat{a} - \langle \boldsymbol \tau_s \rangle  A^{-1} \hat{a} .
\end{aligned} 
\end{equation} 

\section{Proof of Corollary 3.3} When $A=0$ we have the following 
\begin{equation}
e^{ A\tau} =I_n ,  \ \ \     e^{ A\tau  } \int_0^\tau   e^{ -Al} \hat{a} dl  = \tau \hat{a}.
\end{equation}
Further 
\begin{equation}
\frac{1}{\langle \boldsymbol \tau_s\rangle}\left(\int_0^{\infty} e^{-\int_0^\tau h(y) d y}  \right) =\int_0^{\infty} p(\tau)d \tau=1.
\end{equation}
Hence \eqref{mean of x} simplifies to
\begin{equation}
\begin{aligned}
\overline{\langle { {\boldsymbol{x}}} \rangle} = &  J  \left(I_n - J   \right)^{-1}\langle \boldsymbol \tau_s \rangle \hat{a} +   \left(I_n -J   \right)^{-1}\hat{r} + \frac{1}{\langle \boldsymbol \tau_s\rangle}\int_0^{\infty}  \tau   e^{-\int_0^\tau  h(y) d y} d\tau \hat{a} .
\end{aligned}
\label{mean of x1111}	
\end{equation} 

Moreover, from equation \eqref{hr} we can calculate the second-order moment $\langle \boldsymbol \tau_s^{i+1} \rangle $ as
\begin{equation}
{\langle \boldsymbol \tau_s^{i+1} \rangle } = \int_0^{\infty} \tau ^{i+1} h(\tau)e^{-\int_0^{\tau} h(y) d y} d \tau , 
\end{equation}
in which integrating by parts results in 
\begin{equation}
{\langle \boldsymbol \tau_s^{i+1}  \rangle }  =(i+1)\int_0^{\infty}\tau^i e^{-\int_0^{\tau} h(y) d y} d \tau. \label{tau}
\end{equation}
Hence from \eqref{prob. tau} we have the following
\begin{equation}
\langle \boldsymbol \tau^i \rangle = \frac{1}{\langle \boldsymbol \tau_s\rangle}\int_0^{\infty}  \tau^i   e^{-\int_0^\tau  h(y) d y} d\tau  = \frac{{\langle \boldsymbol \tau_s^{i+1} \rangle }}{(i+1){\langle \boldsymbol \tau_s  \rangle }}, \label{mean of tau}
\end{equation}
and by picking $i=1$, \eqref{mean of x1111} simplifies to \eqref{mean of x A=0}. 

\section{Proof of Theorem 3.4}
\subsection{Statistical moments of $ {\boldsymbol \mu }$ after an event}
Based on \eqref{conditional x20}
\begin{equation}
\begin{aligned}
\langle  {\boldsymbol{x}}(\boldsymbol{t}_s^+)  {\boldsymbol{x}}^\top (\boldsymbol{t}_s^+) \rangle = & \langle  {\boldsymbol{x}} (\boldsymbol{t}_s^+)\rangle\langle  {\boldsymbol{x}} (\boldsymbol{t}_s^+)\rangle^\top+Q  {\boldsymbol{x}}(\boldsymbol{t}_s^-) {\boldsymbol{x}}^\top(\boldsymbol{t}_s^-) Q^\top\\ &	
+ B    {\boldsymbol{x}} (\boldsymbol{t}_s^-)\hat{c}^\top+ \hat{c}  {\boldsymbol{x}}^\top  (\boldsymbol{t}_s^-)B^\top+ D.
\end{aligned}
\end{equation}
Further from \eqref{conditional x0}, $\langle  {\boldsymbol{x}} (\boldsymbol{t}_s^+)\rangle\langle  {\boldsymbol{x}} (\boldsymbol{t}_s^+)\rangle^\top  $ can be written as
\begin{equation}\begin{aligned}
\langle  {\boldsymbol{x}} (\boldsymbol{t}_s^+)\rangle\langle  {\boldsymbol{x}} (\boldsymbol{t}_s^+)\rangle^\top   = &  J   {\boldsymbol{x}}(\boldsymbol{t}_s^-) {\boldsymbol{x}}^\top(\boldsymbol{t}_s^-) J^\top\\ &	
+ J    {\boldsymbol{x}} (\boldsymbol{t}_s^-)\hat{r}^\top+ \hat{r}  {\boldsymbol{x}}^\top  (\boldsymbol{t}_s^-)J^\top+ \hat{r}\hat{r}^\top.
\end{aligned}
\end{equation}
Combining these two equations and using \eqref{kronecker} results in \eqref{Jmu}.

\subsection{Necessary and sufficient condition on existence of $ {\boldsymbol \mu }$}
Let us define 
\begin{equation}
 {\boldsymbol u} \equiv  e^{ A\boldsymbol \tau_s} \int_0^{\boldsymbol \tau_s}  e^{ -A l} \hat{a} dl .
\end{equation}
Using \eqref{right before}, the $ {\boldsymbol x}  {\boldsymbol x}^\top$ right before $s^{th}$ event (${ {\boldsymbol{x}}}(\boldsymbol t_{s}^-){ {\boldsymbol{x}}}^\top(\boldsymbol t_{s}^-)$) is related to $ {\boldsymbol x}(\boldsymbol t_{s-1}^+) $ as  
\begin{equation}\begin{aligned}
{ {\boldsymbol{x}}}(\boldsymbol t_{s}^-){ {\boldsymbol{x}}}^\top(\boldsymbol t_{s}^-) = &  { \boldsymbol u} { \boldsymbol u}^\top  +    { \boldsymbol u} \left(  e^{ A\boldsymbol  \tau_s}  {\boldsymbol x}(\boldsymbol t_{s-1}^+)  \right)^\top  
+ \left( e^{ A\boldsymbol \tau_s} {\boldsymbol x}(\boldsymbol t_{s-1}^+)  \right)  { \boldsymbol u}^\top \\ & +  \left( e^{ A\boldsymbol  \tau_s} {\boldsymbol x}(\boldsymbol t_{s-1}^+) \right)  \left( e^{ A\boldsymbol  \tau_s} {\boldsymbol x}(\boldsymbol t_{s-1}^+) \right)^\top.
\end{aligned}
\end{equation}
Thus the mean of the second-order moment of the states after $s^{th}$ event is  
\begin{equation} \begin{aligned}
\langle{ {\boldsymbol{x}}}(\boldsymbol t_{s}^+){ {\boldsymbol{x}}}^\top(\boldsymbol t_{s}^+)  \rangle  =&
Q\left\langle  { \boldsymbol u} { \boldsymbol u}^\top \right \rangle  Q^\top+J\left\langle  { \boldsymbol u} { \boldsymbol u}^\top \right \rangle  J^\top \\ & +Q\left( \left\langle    { \boldsymbol u}  {\boldsymbol x}(\boldsymbol t_{s-1}^+)  ^\top e^{ A^\top \boldsymbol \tau_s}\right \rangle  + \left\langle    { \boldsymbol u}  {\boldsymbol x}(\boldsymbol t_{s-1}^+)  ^\top e^{ A^\top \boldsymbol \tau_s}\right \rangle ^\top  \right)Q^\top  \\
& +J\left( \left\langle    { \boldsymbol u}  {\boldsymbol x}(\boldsymbol t_{s-1}^+)  ^\top e^{ A^\top \boldsymbol \tau_s}\right \rangle  + \left\langle    { \boldsymbol u}  {\boldsymbol x}(\boldsymbol t_{s-1}^+)  ^\top e^{ A^\top \boldsymbol \tau_s}\right \rangle ^\top  \right)J^\top    
\\ & +  Q\left\langle   e^{ A\boldsymbol \tau_s} {\boldsymbol x}(\boldsymbol t_{s-1}^+)  \boldsymbol  x^\top(\boldsymbol t_{s-1}^+)  e^{ A^\top \boldsymbol \tau_s}\right \rangle  Q^\top\\ &
+ J\left\langle   e^{ A\boldsymbol \tau_s} {\boldsymbol x}(\boldsymbol t_{s-1}^+)   \boldsymbol x^\top(\boldsymbol t_{s-1}^+)  e^{ A^\top \boldsymbol \tau_s}\right \rangle  J^\top  + B   \left\langle  { \boldsymbol u}  \right \rangle \hat{c}^\top
\\ & 
+ B   \langle e^{ A\boldsymbol \tau_s}  {\boldsymbol x}(\boldsymbol t_{s-1}^+)  \rangle \hat{c}^\top + J   \left\langle  { \boldsymbol u}  \right \rangle \hat{r}^\top
+ J   \langle e^{ A\boldsymbol \tau_s}   {\boldsymbol x}(\boldsymbol t_{s-1}^+)  \rangle \hat{r}^\top 
\\ &  + \hat{c}   \left\langle  { \boldsymbol u}^\top  \right \rangle B^\top
+ \hat{c}   \langle     {\boldsymbol x}^\top(\boldsymbol t_{s-1}^+)  e^{ A^\top \boldsymbol \tau_s}\rangle B^\top
\\ & + \hat{r} \left\langle  { \boldsymbol u}^\top  \right \rangle J^\top
+ \hat{r}  \langle    {\boldsymbol x}^\top(\boldsymbol t_{s-1}^+) e^{ A^\top \boldsymbol \tau_s}\rangle J^\top + D +\hat{r}\hat{r}^\top  .  
\end{aligned}
\end{equation}
By using vectorization, we have
\begin{equation}\label{second order}  \begin{aligned} 
& {\rm vec}(\langle{ {\boldsymbol{x}}}(\boldsymbol t_{s}^+){ {\boldsymbol{x}}}^\top(\boldsymbol t_{s}^+)  \rangle)  = (J\otimes J+Q\otimes Q)\langle e^{ A\boldsymbol \tau_s}\otimes e^{ A\boldsymbol \tau_s} \rangle {\rm vec}(\langle \boldsymbol x(\boldsymbol t_{s-1}^+)   \boldsymbol x^\top(\boldsymbol t_{s-1}^+ \rangle )\\
& +   (J\otimes J+Q\otimes Q)(\langle e^{ A\boldsymbol \tau_s}\otimes  { \boldsymbol u}\rangle+ \langle  { \boldsymbol u} \otimes e^{ A\boldsymbol \tau_s}\rangle)\langle \boldsymbol x(\boldsymbol t_{s-1}^+ \rangle \\ &  +((B\otimes \hat{c}+J\otimes \hat{r})\langle I_n \otimes e^{ A\boldsymbol \tau_s}\rangle + (\hat{c}\otimes B+\hat{r}\otimes J)\langle e^{ A\boldsymbol \tau_s} \otimes I_n \rangle ) \langle \boldsymbol x(\boldsymbol t_{s-1}^+ \rangle 
\\
&+{\rm vec}( Q\left\langle  { \boldsymbol u} { \boldsymbol u}^\top \right \rangle  Q^\top+J\left\langle  { \boldsymbol u} { \boldsymbol u}^\top \right \rangle  J^\top 
	+ B   \left\langle  { \boldsymbol u}  \right \rangle \hat{c}^\top
	+ J   \left\langle  { \boldsymbol u}  \right \rangle \hat{r}^\top
	\\ & + \hat{c}   \left\langle  { \boldsymbol u}^\top  \right \rangle B^\top
	+ \hat{r} \left\langle  { \boldsymbol u}^\top  \right \rangle J^\top
	+ D +\hat{r}\hat{r}^\top ).
\end{aligned}
\end{equation}
Hence, the steady-state moments of vector $ {\boldsymbol \mu }$ right after an event exists if and only if all the eigenvalues of $(J\otimes J+Q\otimes Q)\left\langle  e^{ A\boldsymbol \tau_s} \otimes e^{ A\boldsymbol \tau_s} \right \rangle$ are inside the unit circle. The rest of the proof is similar to Appendix A.

\section{Proof of Remark 1}
Based on Corollary 11 of \cite{zhz06}, for a negative definite symmetric matrix $M_1$ and a positive semidefinite matrix $M_2$ we have
\begin{equation}
\lambda_{min}(M_1M_2 )\geq\lambda_{min}(M_1) \lambda_{max}( M_2 ),
\end{equation} 
where $\lambda_{min}$ and $ \lambda_{max}$ denote the smallest and largest eigenvalue of a matrix, respectively. 
Based on the fact that exponential of a Hurwitz matrix is positive definite and $-J$ is symmetric negative definite ($J$ is diagonal positive definite) we have 
\begin{equation}
\lambda_{min}(-J \left \langle {\rm e}^{ A \boldsymbol T_s} \right \rangle )\geq\lambda_{min}(-J) \lambda_{max}( {\rm e}^{ A \boldsymbol T_s} ). 
\end{equation}
Given the fact that $\small \lambda_{min}(-J)=-\lambda_{max}(J)$ and $\small \lambda_{min}(-J \left \langle {\rm e}^{ A \boldsymbol T_s} \right \rangle )=-\lambda_{max}(J \left \langle {\rm e}^{ A \boldsymbol T_s} \right \rangle )$, we have 
\begin{equation}
\lambda_{max}(J \left \langle {\rm e}^{ A \boldsymbol T_s} \right \rangle )\leq\lambda_{max}(J) \lambda_{max}( {\rm e}^{ A \boldsymbol T_s} ). \label{remark 1 reason}
\end{equation} 
The proof of the second part of this remark is from the fact that eigenvalues of Kronecker product of two matrices are the multiplication of their eigenvalues \cite{hoj91}.

\section{Extension of Corollary 3.3}
In the limit of $A=0$, $ \langle e^{A_\mu\boldsymbol \tau_s }  \rangle $ in \eqref{mu00} simplifies to
\begin{equation}
A_\mu=  \left[\begin{array}{c;{2pt/2pt}c}
0 & 0\\ \hdashline[2pt/2pt] I_n \otimes \hat{a} + \hat{a}  \otimes I_n & 0
\end{array}\right],  
 \Rightarrow \langle e^{A_\mu\boldsymbol \tau_s }  \rangle  = \left[\begin{array}{c;{2pt/2pt}c}
 I & 0\\ \hdashline[2pt/2pt] (I_n \otimes \hat{a} + \hat{a}  \otimes I_n) \langle \boldsymbol \tau_s \rangle  & I
 \end{array}\right] .
\end{equation}
Moreover 
\begin{equation}
\left \langle  e^{ A_\mu \boldsymbol \tau_s} \int_0^{\boldsymbol \tau_s}  e^{ -A_\mu l} \hat{a}_\mu  dl  \right \rangle = \left[\begin{array}{c}
\hat{a}  \langle \boldsymbol \tau_s \rangle \\ \hdashline[2pt/2pt] \frac{1}{2} (I_n \otimes \hat{a} + \hat{a}  \otimes I_n)\hat{a} \langle \boldsymbol \tau_s^2 \rangle  
\end{array}\right].
\end{equation}
Similarly 
\begin{equation}
\begin{aligned}
& \langle e^{A_\mu\boldsymbol \tau_s }  \rangle  = \left[\begin{array}{c;{2pt/2pt}c}
I & 0\\ \hdashline[2pt/2pt] (I_n \otimes \hat{a} + \hat{a}  \otimes I_n) \langle \boldsymbol \tau \rangle  & I
\end{array}\right], \\
& \left \langle  e^{ A_\mu \boldsymbol \tau_s} \int_0^{\boldsymbol \tau_s}  e^{ -A_\mu l} \hat{a}_\mu  dl  \right \rangle = \left[\begin{array}{c}
\hat{a}  \langle \boldsymbol \tau \rangle \\ \hdashline[2pt/2pt] \frac{1}{2} (I_n \otimes \hat{a} + \hat{a}  \otimes I_n)\hat{a} \langle \boldsymbol \tau^2 \rangle  
\end{array}\right].
\end{aligned}
\end{equation}
By using \eqref{mean of tau} we see that all the terms in \eqref{mu00} are only depending on the first three moments of $\boldsymbol \tau_s$. 

\section{Matrices needed to calculate $\overline{\langle  {\boldsymbol \mu} \rangle }$ for a gene product}
\begin{equation}
\begin{aligned}
&   \langle e^{A_\mu\boldsymbol \tau_s } \rangle  = \left[\begin{array}{cc}
\langle e^{-\gamma_x \boldsymbol \tau_s}\rangle & 0\\ 2\frac{k_x}{\gamma_x}\left(\langle e^{-\gamma_x \boldsymbol \tau_s}\rangle + \langle e^{-2\gamma_x \boldsymbol \tau_s}\rangle \right) &   \langle e^{-2\gamma_x \boldsymbol \tau_s }\rangle 
\end{array}\right]  ,\\
&\left \langle e^{ A_\mu\boldsymbol \tau_s } \int_0^{\boldsymbol \tau_s }   e^{ -A_\mu l} \hat{a}_\mu dl \right \rangle = 
\left[\begin{array}{c}
\frac{k_x}{\gamma_x } (1-\langle e^{-\gamma_x \boldsymbol \tau_s}\rangle) \\ \frac{k^2 }{\gamma_x^2 } (\left\langle e^{ -2\gamma_x \boldsymbol \tau_s}  \right \rangle -2 \left\langle e^{ -\gamma_x \boldsymbol \tau_s }  \right \rangle +1)
\end{array}\right],\label{matrices1}
\end{aligned} 
\end{equation}
and 
\begin{equation}
\begin{aligned}
&   \langle e^{A_\mu\boldsymbol \tau} \rangle  = \left[\begin{array}{cc}
\langle e^{-\gamma_x \boldsymbol \tau}\rangle & 0\\ 2\frac{k_x}{\gamma_x}\left(\langle e^{-\gamma_x \boldsymbol \tau}\rangle + \langle e^{-2\gamma_x \boldsymbol \tau}\rangle \right) &   \langle e^{-2\gamma_x \boldsymbol \tau}\rangle 
\end{array}\right]  ,\\
&\left \langle e^{ A_\mu\boldsymbol \tau  } \int_0^{\boldsymbol \tau}   e^{ -A_\mu l} \hat{a}_\mu dl \right \rangle = 
\left[\begin{array}{c}
\frac{k_x}{\gamma_x } (1-\langle e^{-\gamma_x \boldsymbol \tau}\rangle) \\ \frac{k^2 }{\gamma_x^2 } (\left\langle e^{ -2\gamma_x \boldsymbol \tau}  \right \rangle -2 \left\langle e^{ -\gamma_x \boldsymbol  \tau }  \right \rangle +1)
\end{array}\right] .\label{matrices2}
\end{aligned} 
\end{equation}
Using the fact that
\begin{equation}
\langle e^{-\gamma_x\boldsymbol \tau} \rangle  =\frac{1}{\langle \boldsymbol \tau_s \rangle }\frac{1}{\gamma_x}\left(  1- \left\langle e^{ -\gamma_x \boldsymbol  \tau_s}  \right \rangle \right), \ \ \  \langle e^{-2\gamma_x\boldsymbol \tau} \rangle  =\frac{1}{\langle \boldsymbol \tau_s \rangle }\frac{1}{2\gamma_x}\left(  1- \left\langle e^{ -2\gamma_x \boldsymbol  \tau_s}  \right \rangle \right) \label{T is}
\end{equation}
(See Appendix C), equation \eqref{matrices2} can be changed to just contain expected values with respect to $\boldsymbol \tau_s$. Putting these matrices and vectors back in Theorem 3.4 we derive $\overline{\langle \boldsymbol x^ \rangle}$ as in \eqref{second order00}.

\section{Proof of Theorem 5.1}
	Using \eqref{dynamics}, the states of TTSHS right before $s^{th}$ event is 
	\begin{equation}\begin{aligned}
	 {\boldsymbol{x}}(\boldsymbol{t}_s^-) =\int_{0}^{\boldsymbol \tau_s}\varPhi(\boldsymbol \tau_s,l)\hat{a}(l)dl 
	+\varPhi(\boldsymbol \tau_s,0)  {\boldsymbol{x}}(\boldsymbol{t}_{s-1}^+) .  
	\end{aligned}
	\end{equation}  
Thus, the mean of the states after $s^{th}$ event is  
\begin{equation}\begin{aligned}
\langle { {\boldsymbol{x}}}(\boldsymbol t_{s}^+)   \rangle = \left \langle J(\boldsymbol \tau_s)\int_{0}^{\boldsymbol \tau_s}\varPhi(\boldsymbol \tau_s,l)\hat{a}(l)dl    \right \rangle 
+ \left \langle J(\boldsymbol \tau_s )  \varPhi(\boldsymbol \tau_s,0)\right \rangle  \left \langle  { {\boldsymbol{x}}}(\boldsymbol t_{s-1}^+)   \right \rangle  +\langle \hat{r}(\boldsymbol \tau_s ) \rangle . 
\end{aligned}
\end{equation} 
Hence, for steady-state mean of states to not blow up, eigenvalues of $ \left\langle J(\boldsymbol \tau_s) \varPhi(\boldsymbol \tau_s,0)  \right \rangle$ should be inside the unit circle. In this limit the mean of the states just after an event in steady-state ($s\rightarrow \infty$) is
	\begin{equation}
	\begin{aligned}
\lim_{s\to\infty}\langle { {\boldsymbol{x}}}(\boldsymbol t_{s}^+)   \rangle = &\left(I_n - \left \langle  J(\boldsymbol \tau_s) \varPhi(\boldsymbol \tau_s,0)  \right  \rangle  \right)^{-1} \left(\left \langle J(\boldsymbol \tau_s)\int_{0}^{\boldsymbol \tau_s}\varPhi(\boldsymbol \tau_s,l)\hat{a}(l)dl    \right \rangle  + \langle \hat{r}(\boldsymbol \tau_s) \rangle  \right) . \label{xi2} 
	\end{aligned}
	\end{equation}
	By using equation \eqref{xi2}, the steady-state mean of the states in between events for any $\boldsymbol \tau $ is  
	\begin{equation}
	\begin{aligned}
 \lim_{s \rightarrow \infty}&\langle { {\boldsymbol{x}}}(\boldsymbol t_{s}+\boldsymbol \tau)  \rangle=  \varPhi(\boldsymbol \tau,0) \left(I_n - \left \langle  J(\boldsymbol \tau_s) \varPhi(\boldsymbol \tau_s,0)  \right  \rangle  \right)^{-1} \times \\ &  \left(\left \langle J(\boldsymbol \tau_s)\int_{0}^{\boldsymbol \tau_s}\varPhi(\boldsymbol \tau_s,l)\hat{a}(l)dl    \right \rangle  + \langle \hat{r}(\boldsymbol \tau_s) \rangle  \right) +  \int_{0}^{\boldsymbol \tau}\varPhi(\boldsymbol \tau,l)\hat{a}(l)dl    . \label{xi23} 
	\end{aligned}
	\end{equation}
Thus, taking the expected value of $\langle { {\boldsymbol{x}}}(\boldsymbol t_{s}+\boldsymbol \tau)  \rangle$ with respect to $\boldsymbol \tau$ results in the mean of the states as in \eqref{mean of xx}. The rest of proof is similar to that of Theorem 3.1.

\bibliographystyle{IEEEtran}
\bibliography{RefMaster}
\end{document}